\documentclass[acmsmall,screen,review=false,nonacm]{acmart}
\usepackage{algorithm}
\usepackage{algorithmicx}
\usepackage{algpseudocode}

\usepackage{color,soul}
\usepackage{proof}
\usepackage{amsmath}
\usepackage{algorithm}%
\usepackage{algpseudocode}%
\usepackage{amsthm}
\usepackage{xspace}
\usepackage{wrapfig}
\usepackage{listings}

\usepackage{array}

\newcommand{\setof}[1]{\ensuremath{\left \{ #1 \right \}}}
\newcommand{\tuple}[1]{\ensuremath{\left \langle #1 \right \rangle }}

\newcommand{\variables}[1]{\ensuremath{\mathbf{vars}(#1)}}

\newcommand{\dom}[1]{\ensuremath{\mathbf{dom}\;#1}}
\newcommand{\buildit}{BuildIt\xspace}
\newcommand{\dyn}[1]{\myinline{dyn\_var<#1>}\xspace}
\newcommand{\static}[1]{\myinline{static\_var<#1>}\xspace}

\newcommand{\prophecy}[1]{\myinline{prophecy\_var<#1>}\xspace}

\newcommand{\bldit}[1]{\myinline{#1}}
\newcommand{\blditsmall}[1]{{\footnotesize \lstinline{#1}}}

\newcolumntype{L}[1]{>{\raggedright\let\newline\\\arraybackslash\hspace{0pt}}p{#1}}
\newcolumntype{C}[1]{>{\centering\let\newline\\\arraybackslash\hspace{0pt}}p{#1}}
\newcolumntype{R}[1]{>{\raggedleft\let\newline\\\arraybackslash\hspace{0pt}}p{#1}}

\newcommand{\punt}[1]{}

\let\comment=\relax
\newcommand{\comment}[1]{}
\ifthenelse{0=1} {
\newcommand{\ajay}[1]{{\color{blue} {\bf Ajay:} #1}}
\newcommand{\saman}[1]{{\color{magenta} {\bf Saman:} #1}}
\newcommand{\martin}[1]{{\color{red} {\bf Martin:} #1}}
} {
\newcommand{\ajay}[1]{}
\newcommand{\saman}[1]{}
\newcommand{\martin}[1]{}
}

\setcopyright{none}

\usepackage{booktabs}   %
\usepackage{subcaption} %
\def\OPTL{\textrm{$[$}}
\def\OPTR{\textrm{$]$}}

\definecolor{lightbackground}{rgb}{.98,.98,.97}
\definecolor{lightishbackground}{rgb}{.98,.98,.98}

\definecolor{darkgray}{rgb}{.4,.4,.4}
\definecolor{lightgray}{rgb}{.6,.6,.6}
\definecolor{darkred}{rgb}{.6,0,0}
\definecolor{darkgreen}{rgb}{0.3,.8,.3}
\definecolor{darkdarkgreen}{rgb}{0.0,.7,.0}

\definecolor{darkblue}{rgb}{0,0,.6}
\lstdefinelanguage{default}{
  basicstyle=\small\ttfamily,
}

\definecolor{darkgraycode}{rgb}{0.25,0.25,0.25}
\definecolor{keywordcolor}{RGB}{255,75,175}
\lstdefinelanguage{buildit}{%
  keywords={[1]},%
  keywords={[2]void,char,short,long,int,float,double,boolean,size_t,int32_t,delete,bool,if,while,for,else,template,typename,return,struct,class,goto,enum,virtual,public,auto,new,const, unsigned, include, using, namespace, operator, explicit, try, catch, throw, typedef, private, static_cast, override, constexpr, consteval, Int, Float, then, true, false, static},
  keywords={[3]dyn_var,static_var, union_var, prophecy_var, prophecy, builder, block, builder_context},
  keywords={[4]},
  literate={[OPT[}{{\OPTL}}1 {]OPT]}{{\OPTR}}1,
  string=[b]",
  comment=[l]//,
  morecomment=[s]{/*}{*/},
  mathescape=true,
  flexiblecolumns=true,
  tabsize=2,
  captionpos=b,
  basicstyle=\scriptsize\ttfamily,
  keywordstyle={[1]\color{darkred}\bfseries},
  keywordstyle={[2]\color{blue}\bfseries},
  keywordstyle={[3]\color{darkdarkgreen}\bfseries},
  keywordstyle={[4]\color{darkred}\bfseries},
  numbers=left,
  stepnumber=1,
  numberstyle=\scriptsize\color{black}\ttfamily,
  emphstyle=\slshape,
  commentstyle=\color{darkgray},
  stringstyle=\color{darkdarkgreen},
  frame=ltbr,
  keepspaces=true,
  showstringspaces=false,
  backgroundcolor=\color{lightbackground},
  aboveskip=0pt,
  belowskip=0pt,
  lineskip=0pt,
  xleftmargin=2em,
}

\lstdefinelanguage{mypython}{
  basicstyle=\scriptsize\ttfamily,
  keywords={[1]},%
  keywords={[2]if,while, for,else,return,class,def,import,as,lambda, Float, Int, Double},
  keywords={[3]},
  keywords={[4]},
  literate={[OPT[}{{\OPTL}}1 {]OPT]}{{\OPTR}}1,
  string=[b]",
  comment=[l]{\#},
  morecomment=[s]{/*}{*/},
  mathescape=true,
  flexiblecolumns=true,
  tabsize=2,
  captionpos=b,
  keywordstyle={[1]\color{darkred}},
  keywordstyle={[2]\color{blue}},
  keywordstyle={[3]\color{darkdarkgreen}\bfseries},
  keywordstyle={[4]\color{darkred}\bfseries},
  numbers=left,
  stepnumber=1,
  numberstyle=\scriptsize\color{black}\ttfamily,
  emphstyle=\slshape,
  commentstyle=\color{darkgray},
  stringstyle=\color{darkdarkgreen},
  frame=ltbr,
  keepspaces=true,
  backgroundcolor=\color{lightbackground},
  xleftmargin=2em,
}

\lstdefinelanguage{graphit}{
  keywords={[1]},%
  keywords={[2]func, end, apply, VertexSet, EdgeSet, from, to, Vertex},
  keywords={[3]},
  keywords={[4]},
  literate={[OPT[}{{\OPTL}}1 {]OPT]}{{\OPTR}}1,
  string=[b]",
  comment=[l]{\#},
  morecomment=[s]{/*}{*/},
  mathescape=true,
  flexiblecolumns=true,
  tabsize=2,
  captionpos=b,
  basicstyle=\footnotesize\ttfamily,
  keywordstyle={[1]\color{darkred}},
  keywordstyle={[2]\color{blue}},
  keywordstyle={[3]\color{darkdarkgreen}\bfseries},
  keywordstyle={[4]\color{darkred}\bfseries},
  numbers=left,
  stepnumber=1,
  numberstyle=\scriptsize\color{black}\ttfamily,
  emphstyle=\slshape,
  commentstyle=\color{darkgray},
  stringstyle=\color{darkdarkgreen},
  frame=ltbr,
  keepspaces=true,
  xleftmargin=2em,
}

\newcommand{\lstreset}{\lstset{language=buildit,backgroundcolor=\color{darkgray}, belowskip=.5em, aboveskip=.5em}}
\lstreset

\newcommand{\titletext}{Prophecy Variables, Program Analysis, Binding Times, and Staged Compilation in the BuildIt System}

\begin{document}	
	\title[\titletext]{\titletext}

    \author{Ajay Brahmakshatriya}
\affiliation{
  \institution{MIT EECS and MIT CSAIL}
  \city{Cambridge}
  \country{USA}
}
\email{ajaybr@mit.edu}

\author{Saman Amarasinghe}
\affiliation{
  \institution{MIT EECS and MIT CSAIL}
  \city{Cambridge}
  \country{USA}
}
\email{saman@csail.mit.edu}

\author{Martin Rinard}
\affiliation{
  \institution{MIT EECS and MIT CSAIL}
  \city{Cambridge}
  \country{USA}
}
\affiliation{
  \institution{NUS}
  \country{Singapore}
}
\email{rinard@csail.mit.edu}

\renewcommand{\shortauthors}{Brahmakshatriya, Amarasinghe, and Rinard}

	\begin{abstract} 

\buildit is a lightweight system for implementing domain specific languages and compilers using multi-stage execution. By eliminating the need to implement components such as parsers, intermediate representations, and code generators, \buildit can significantly reduce the engineering effort and compiler expertise required to implement DSLs. \buildit is implemented as a collection of C++ libraries that use partial evaluation to generate optimized DSL code. These libraries use history variables and standard forward C++ program execution to obtain analysis results traditionally obtained by forward
dataflow analyses. However, a major drawback of a system that relies purely on standard forward execution and lacks support for an IR, is that analyses that propagate information backwards are impossible to implement. These include analyses that enable critical optimizations such as optimizing data-copy between devices based on uses, operator fusion, and binding-time analyses. 

To address these limitations, we introduce a novel technique that uses prophecy variables to predict information about the future of the program and
apply this information to optimize the generated code. The \buildit multi-stage implementation uses a repeated first stage program execution mechanism to ensure that the prophecy variable predictions are accurate, with repeated first stage execution times amortized over subsequent second stage executions.

We show that such a prophecy variable based implementation can deliver analysis results that would typically require backwards dataflow analyses over an explicit IR while preserving the fundamental simplicity of \buildit. Because this implementation leverages the existing \buildit reexecution machinery, it adds little overall complexity to the existing \buildit implementation. We use prophecy variables to implement three analyses and corresponding optimizations and demonstrate that these optimizations can improve the performance up to 5.5x while also significantly simplifying the programming interface.

\end{abstract}

	\maketitle
	
	\pagestyle{fancy}
\fancyfoot{}
\section{Introduction}

Over the last decade, domain specific languages (DSLs) have become a vehicle of choice for high-performance computations. Focusing on a specific domain not only enhances programmability by providing domain concepts directly in the programming language but also enhances performance by enabling optimizations and transformations that leverage common patterns in the domain. In contrast, libraries that expose a set of operations typically confine their optimizations within a single operator. Since DSL compilers have a complete view of the program written by the end user, they can analyze and transform across a sequence of operations to generate highly optimized code. Examples of such optimizations in the tensor domain include efficiently managing data placement and movement across the full multi-operation computation (as opposed to within a single operation), fusing sequences of operations into a single efficient integrated computation, and simplifying and specializing operators to specific values that occur at run time.

A common issue, however, is implementation overhead and complexity. DSL compiler implementations often involve
parsers, %
semantic checkers, and program analyses over a range of intermediate representations. The resulting implementation effort can require multiple tens of thousands of lines of code (in some cases over a hundred thousand; see Table~\ref{tbl:dsllocperf}), posing a substantial obstacle to DSL development and deployment. Beyond implementation effort, DSL development often requires compiler experts: although domain experts understand the DSL abstraction and domain-specific optimizations, most lack the specialized programming language implementation expertise required to build highly optimized DSL implementations.

\begin{table}[tbp]
    \centering
    \small
    \begin{tabular}{|l|r|r|r|}
        \hline
        Standard DSL Implementation vs. & \multicolumn{2}{c|}{Implementation Lines of Code} & BuildIt Speedup \\
        BuildIt  & Standard DSL & BuildIt & Geometric Mean Speedup \\ \hline
        \hline
        GraphIt \cite{g2} / GraphIt (BuildIt)\cite{buildsl} & 23,783 & 2,021 & 3\% \\ \hline
        COLA \cite{shim} / SHIM (BuildIt) \cite{shim} & 7,500 & 5,700 & 1.4\%  \\ \hline
        PCRE2\cite{pcre2} / BREze (BuildIt) \cite{tamara:meng-thesis:2023} & 133,995 & 1,564 & 800\%  \\ \hline
        RE2\cite{re2} / BREze (BuildIt) \cite{tamara:meng-thesis:2023} & 26,587 & 1,564 &  37\% \\ \hline
        Hyperscan\cite{hyperscan} / BREze (BuildIt) \cite{tamara:meng-thesis:2023} & 187,033 & 1,564 & -127\% \\ \hline
    \end{tabular}
 \vspace{.1in}
    \caption{\label{tbl:dsllocperf}Implementation effort and performance of DSLs implemented as standalone compilers or libraries versus corresponding \buildit implementations. The second and third columns show implementation lines of code for the standard and \buildit versions, respectively. The last column shows the geometric mean relative speedup of \buildit over the baseline on benchmark suites common to the domain. A negative value indicates that \buildit is slower.}
\center
\small
\begin{tabular}{|l|r|r|}
\hline
Component & BuildIt\cite{buildsl} & GraphIt\cite{g2} \\ \hline \hline
Frontend & - & 9,593 \\ \hline
Scheduling Language & 151 &  2,401 \\ \hline
Midend Analysis and Transformations & - & 9,601 \\ \hline
Types and Operator Implementation & 1,320 & - \\ \hline
Code Generation & 550 & 2,188 \\ \hline
{\bf Total} & {\bf 2,021} & {\bf 23,783} \\ \hline
\end{tabular}
\vspace{.1in}
\caption{\label{tbl:breakdown}Implementation lines of code by component for GraphIt as a standalone DSL compiler versus its \buildit implementation. The first column lists compiler components; the remaining columns give implementation lines of code for \buildit and GraphIt, respectively. Both implementations provide the same functionality and optimizations. \buildit requires an order of magnitude less code while maintaining the same performance.}
\label{tab:linesofcode}
\vspace{-.2in}
\end{table}

\subsection{BuildIt}

In response to this situation, BuildIt~\cite{buildit,ajay-thesis, buildsl} leverages modern programming language features such as partial evaluation and multi-stage execution to implement DSLs. Operating completely within a modern general purpose language (C++), BuildIt implements compilation and optimization techniques that operate across the entire domain-specific multi-operation computation. The \buildit framework divides execution into two distinct phases, each capable of performing arbitrary C++ computations. During the first stage, the available values are completely evaluated, and control flow constructs such as conditions and loops over those values are resolved. The output of the first stage is a program containing operations on second-stage values specialized to the first-stage execution. Thus, the first-stage logic has complete control over the generated second-stage program, making it possible to implement analyses and optimizations in a natural way. This use of first-stage logic to specialize the generated code eliminates the need for parsers, explicit intermediate representations, or dataflow analyses and transformations over such IRs, greatly reducing the lines of code required and the programming language expertise expected of the developer.

Table~\ref{tbl:dsllocperf} quantifies this reduction by counting lines of code in corresponding BuildIt and standard DSL implementations. Table~\ref{tbl:breakdown} breaks down the lines of code for the BuildIt and GraphIt implementations (both provide the same functionality and implement the same optimizations).
These numbers show that \buildit implementations can require up to 100 times fewer lines of code than standard implementations of the same DSL, a reduction substantial enough to transform the viability of many DSLs. The performance results in Table~\ref{tbl:dsllocperf} also show that BuildIt implementations often exhibit comparable performance~\cite{buildsl,ajay-thesis} while requiring a fraction of the implementation effort. The only case where BuildIt is slower is in comparison to Hyperscan~\cite{hyperscan}, which is at least 100 times larger and includes a complex regular expression library with hand-optimized SIMD and other vector instructions.

\subsection{Forward Program Analysis in BuildIt}
Efficient DSL compiler implementations require analyses that span multiple operators and parameterized optimizations. \buildit, implemented as a C++ library, uses standard forward program execution for information collection and data propagation in the first stage. Forward execution allows passing information from one operator to subsequent operators. This supports forward analyses that propagate information about past execution to optimization sites, specifically via {\em history variables} (first-stage variables)~\cite{OwickiG76, AbadiL91, rinard2020dataflow}.
History variables are assigned as the \buildit first stage executes and carry analysis information that can be accessed later, still during the first stage, to optimize the generated second-stage code. In this way, \buildit implements optimizations that traditional compilers instead achieve through forward dataflow analyses over explicit intermediate representations. Because this information is carried and computed entirely in the first stage, second-stage execution incurs no overhead. Examples include constant propagation, reaching definitions, and shared memory access optimizations~\cite{buildsl}.

\subsection{Prophecy Variables in BuildIt}
But forward program execution and history variables cannot enable optimizations that require information about the future execution of the program. This information is needed for optimizations such as liveness, operator fusion, and binding time analysis. Traditional compilers obtain it through backward program analyses over explicit intermediate representations~\cite{DragonBook,appel2004modern,muchnick1997advanced,cooper2011engineering,kennedy2001optimizing}.

In this paper, we introduce prophecy variables, which use first-stage angelic nondeterminism to predict properties of operators that appear later in the program. Subsequent prophecy variable checks filter out incorrect predictions. Prophecy variables were initially developed to prove forward simulation relations in state machine refinement proofs involving reasoning about future program behavior~\cite{AbadiL91}. We show that, in multi-stage DSL compiler implementations such as \buildit, prophecy variables can obtain analysis results that would typically require backward dataflow analyses over an intermediate representation. By introducing this new construct, we eliminate a major drawback of \buildit compared to traditional DSL compilers, allowing DSL designers to reason about both past and future execution while retaining the ability to implement sophisticated optimizations without an explicit intermediate representation.

We also present an implementation of prophecy variables in \buildit as a C++ library. This implementation uses a reexecution based approach to obtain correct predictions, leveraging the existing reexecution machinery in \buildit for exploring dynamic control flow. Since prophecy variables and the reexecutions associated with them are confined to the first stage, while most \buildit programs perform the bulk of performance-sensitive work in the second stage, this approach adds very little overhead, which can be amortized over repeated executions of the generated program.

\subsection{Contributions}

This paper makes the following contributions:
\begin{itemize}

    \item {\bf Program Analysis Technique to Reason About the Future}: We present a novel program analysis technique using prophecy variables to obtain information about future program execution without relying on concrete intermediate representations. This addresses a major limitation of multi-stage DSL construction in systems like \buildit, enabling optimizations that require future execution information while preserving BuildIt's lightweight DSL implementation approach and substantially reducing compiler implementation effort.

    \item {\bf Prophecy Variables in \buildit}: We present an algorithm for implementing prophecy variables in standard programming languages with forward execution by predicting their values and re-executing upon mispredictions. We implement this mechanism in \buildit, leveraging its existing reexecution machinery for exploring dynamic control flow. Multi-stage execution confines the overhead of prophecy variables and reexecution to the first stage, allowing it to be amortized. Prophecy and history variables are encapsulated within the BuildIt data structures implementing the DSL, requiring little to no knowledge of these concepts from application developers.

    \item {\bf Formalization of Prophecy Variables}: We formalize prophecy variables and the interaction of their reexecution mechanism with the existing two-stage execution in systems like \buildit. We also present theorems mechanized in Lean~4 establishing the correctness of the first-stage reexecution mechanism.

    \item {\bf Implementation of Data-flow Analysis using Prophecy Variables}: We implement three program analyses that reason about future execution for performance optimizations: i) CPU/GPU data movement optimizations, ii) Convolution and ReLU operator fusion in a machine learning DSL, and iii) Binding time analysis enabling static/dynamic variables in \buildit. The last enables a new form of binding time polymorphism, automatically specializing reusable abstractions to static or dynamic variables based on whole-program usage, further simplifying high-performance DSL development in \buildit.

    \item {\bf Performance Results}: We evaluate the performance impact of the first two analyses and their transformations, demonstrating improvements of up to 5.5x and 1.1x, respectively, while measuring the first-stage execution overhead of prophecy variables. For binding time analysis, we measure the overhead of a simplified \buildit interface that eliminates the need for users to annotate variables as static (first stage) or dynamic (second stage).

\end{itemize}
The remainder of the paper is structured as follows. Section~\ref{sec:example} presents an example that illustrates the use of prophecy variables in BuildIt. 
Section~\ref{sec:formalization} presents a formal treatment
of prophecy variables and correct prophecy variable prediction
via program reexecution in the context of a core programming language. Section~\ref{sec:applications} discusses the implementation of prophecy variables in the BuildIt system and presents performance and engineering effort results for benchmark BuildIt computations that use optimizations enabled by prophecy variables. We discuss related work in Section~\ref{sec:related} and conclude in Section~\ref{sec:conclusion}. 

    \section{Example}
\label{sec:example}
\newcommand{\myinline}[1]{\lstinline[basicstyle={\ttfamily\normalsize}]{#1}}

\begin{figure}
\begin{lstlisting}[language=buildit]
namespace runtime {
dyn_var<int(const char*, int)> open_file;
dyn_var<int(int)> read_byte;
dyn_var<void(int, int)> write_byte;
dyn_var<void(int)> close_handle;
}

void process_file (dyn_var<const char*> path) {
    // TrueTop describes a boolean lattice with True at the Top. 
    prophecy_var<TrueTop> needs_write(TrueTop::FALSE); 
    
    // Open handle in write mode ONLY if a write access is required later
    dyn_var<int> handle;
    if (needs_write.is_at_most(TrueTop::FALSE)) {
        handle = runtime::open_file(path, ios::read);
    } else {
        handle = runtime::open_file(path, ios::read | ios::write);
    }

    // Read operation does not constrain the prophecy var
    dyn_var<int> byte = runtime::read_byte(handle);

    // Write operation requires the lattice value to be at least TRUE
    needs_write.assert_requires_atleast(TrueTop::TRUE);
    runtime::write_byte(handle, byte);

    runtime::close_handle(handle); 
} 
\end{lstlisting}

    \caption{Prophecy variables predict FILE access modes.}
    \label{fig:filehandlepredict}
\end{figure}

Figure~\ref{fig:filehandlepredict} presents an example for managing file access modes based on types of access later in the application written in C++. The example uses prophecy variables to identify if the file handle will be used to perform a write access and conditionally opens the file handle as read only or read and write. The application demonstrates the use of prophecy variables in the multi-stage framework \buildit explained in Section~\ref{sec:buildit}. The relevant \buildit type here is \dyn{T} which is used to declare variables and expressions that are generated to be executed in the second stage. Since the handle and the runtime functions are all declared to be of type \dyn{T}, all the actual file handling code is generated. The prophecy variable predictions and specialization are performed in the first stage. 

The example declares \texttt{needs\_write} with type \prophecy{TrueTop}, a two point boolean lattice with \texttt{TRUE} at the top and \texttt{FALSE} at the bottom. The prophecy variable uses angelic non-determinism to predict if the handle will need write accesses in the future. The default prediction is  \texttt{FALSE} (the most optimistic choice if no write constraints are seen in the future), but the prophecy variable can take on any value required to pass all of the prophecy variable checks. The program next checks if \texttt{needs\_write} is at most (less than equal to) \texttt{FALSE} in the lattice. In that case it opens the file handle as read-only else if the variable is higher in the lattice (\texttt{TRUE}), it opens the file handle as read-write. The rest of the code in the application comes from a different part of the abstraction (user code) but has been shown together in the same function for brevity. The user can perform read operations like \texttt{read\_byte} freely without any assertions on the prophecy variable. However, when performing the write operation, the program checks that \texttt{needs\_write} is at least \texttt{TRUE}. 
This check filters out predictions that do not enable writes. Specifically, the initial prediction of \texttt{FALSE} causes the first stage to generate a read-only file open operation. When the first stage encounters the assertion before the write operation, the prediction is found to be incorrect and updated to \texttt{TRUE}, triggering a reexecution. On reexecution, the first stage generates a read-write file open operation, and the prophecy variable assertion succeeds.

This example demonstrates several key prophecy variable constructs, specifically \\\texttt{assert\_requires\_atleast}, which requires the prophecy variable value to be higher than or equal to the parameter in the lattice, and \texttt{is\_at\_most}, which checks if the prophecy variable prediction is lower than or equal to the parameter in the lattice and branches accordingly. Prophecy variables also have a \texttt{get} operator that returns the predicted value (not shown in this example). In classic angelic non-deterministic fashion, only those program executions are valid where all the prophecy variable assertions return true.

	\section{Prophecy Variable Formalization}
\label{sec:formalization}

\newcommand{\VP}{V_{P}}
\newcommand{\VS}{V_{S}}
\newcommand{\VD}{V_{D}}
\newcommand{\kw}[1]{\mathbf{#1}}
\newcommand{\cassign}[3]{#1 := #2;\;\kw{goto}\;#3}
\newcommand{\cifz}[3]{\kw{ifz}\;#1\;\kw{goto}\;#2\;\kw{else}\;#3}
\newcommand{\cgoto}[1]{\kw{goto}\;#1}
\newcommand{\ccheck}[3]{\kw{check}\;#2 \leq #1;\;\kw{goto}\;#3}
\newcommand{\cbelow}[4]{\kw{if}\;#2 \leq #1\;\kw{goto}\;#3\;\kw{else}\;#4}
\newcommand{\cget}[3]{#1 := \kw{get}\;#2;\;\kw{goto}\;#3}
\newcommand{\fail}{\mathbf{fail}}
\newcommand{\den}[1]{[\![#1]\!]}
\newcommand{\Lab}{L}
\newcommand{\Int}{\mathbb{Z}_{64}}
\newcommand{\Nat}{\mathbb{N}}
\newcommand{\fails}{\mathbf{mispredicted}}
\newcommand{\pfun}{\rightharpoonup}
\newcommand{\Fail}{F}
\newcommand{\var}{\kw{var}\;}
\newcommand{\Succ}[1]{\ensuremath{\mathbf{succ}\;#1}}
\newcommand{\Prog}{\mathit{Prog}}
\newcommand{\IfThen}[2]{\State \algorithmicif\ #1\ \algorithmicthen\ #2}
\algdef{C}[IF]{IF}{ElsIfLine}[2]{\algorithmicelse\ \algorithmicif\ #1\ \algorithmicthen\ #2}
\newcommand{\ElseLine}[1]{\State \algorithmicelse\ #1}
\newcommand{\leanname}[2]{{\noindent\raggedright\textit{Lean:} #1 (\texttt{#2}).\par}}

We next formalize prophecy variables, including the interaction between the prophecy variable
reexecution mechanism and the BuildIt dynamic control flow exploration algorithm. 
The formalism works with a core language that includes dynamic, static, and prophecy variables.
We have mechanized the formalism, including theorems that characterize relevant first stage and
second stage correctness properties, in Lean~4.

\subsection{Core Language}
\label{sec:language}

A program $P = \tuple{\VP, \VS, \VD, l_0, M, O}$ has prophecy variables $\VP$, static variables
$\VS$, dynamic variables $\VD$, an entry label $l_0$, a finite map $M : \Lab \pfun C$ from labels
${\Lab = \setof{l_0, l_1, \ldots}}$ to commands (Figure~\ref{fig:corelanguage}), and observable
variables $O \subseteq \VD$. We write $P(l)$ for $M(l)$ and $\Prog$ for the set of programs. Static
and dynamic variables hold 64-bit integers $n \in \Int$, the two's-complement integers
$-2^{63} \leq n < 2^{63}$ with wraparound arithmetic. Each prophecy variable $p$ ranges over a lattice
$\tuple{L_p, \leq_p}$ with bottom $\bot_p$ and join $\vee_p$. As
in CompCert~\cite{Leroy09JAR}, a recent verified compiler~\cite{Nexis}, and the basic block
terminators of LLVM~\cite{LattnerA04}, each command names its successors:
\[
\begin{array}{r@{\;=\;}l@{\qquad\qquad}r@{\;=\;}l}
\Succ{(\cassign{x}{e}{g})} & \setof{g} & \Succ{(\ccheck{p}{r}{g})} & \setof{g} \\
\Succ{(\cifz{x}{g_z}{g_{nz}})} & \setof{g_z, g_{nz}} & \Succ{(\cbelow{p}{r}{g_t}{g_f})} & \setof{g_t, g_f} \\
\Succ{(\cgoto{g})} & \setof{g} & \Succ{\kw{halt}} & \emptyset \\
\multicolumn{2}{c}{} & \Succ{(\cget{x}{p}{g})} & \setof{g}
\end{array}
\]
There are three prophecy commands. A check 
$\ccheck{p}{r}{g}$ command corresponds to the \buildit \texttt{assert\_requires\_atleast}
construct and requires the prophecy variable $p$ to have a value at least $r$, where 
$r$ is either a value $u \in L_p$ or a prophecy variable $q$ with the same lattice as $p$.
A prophecy branch command $\cbelow{p}{r}{g_t}{g_f}$ corresponds to the \buildit \texttt{is\_at\_most}
construct and checks the value of $p$ against $r$ and branches accordingly. 
A $\cget{x}{p}{g}$ command copies the
value of $p$ into $x$; to support $\kw{get}$, the formalization works with lattices whose 
elements can be represented as nonnegative 64-bit integers. For a
function $f$ defined on $\Succ{c}$, $c[f]$ is $c$ with each successor $g$ replaced by $f(g)$.
$\variables{e}$ is the set of variables that $e$ reads.

\begin{definition}[Static independence]
\label{def:staticind}
$P$ is \emph{statically independent} if $\VS \cap \VD = \emptyset$ and $\variables{e} \subseteq \VS$ whenever $P(l) = \cassign{x}{e}{g}$ with
$x \in \VS$.
\end{definition}

\begin{figure}[tp]
\[
\begin{array}{rcl}
a & ::= & n \mid x \\
e \in E & ::= & a \mid \ominus\, a \mid a_0 \oplus a_1 \\
r & ::= & u \mid q \\
c \in C & ::= & \cassign{x}{e}{g} \mid \cifz{x}{g_z}{g_{nz}} \mid \cgoto{g} \mid \kw{halt} \\
  & \mid & \ccheck{p}{r}{g} \mid \cbelow{p}{r}{g_t}{g_f} \mid \cget{x}{p}{g}
\end{array}
\]
\caption{\label{fig:corelanguage} Core Language. $n \in \Int$; $x \in \VS \cup \VD$; $a$ is an atom;
$\ominus$ and $\oplus$ are unary and binary integer operators; $g, g_z, g_{nz}, g_t, g_f \in \Lab$;
$p, q \in \VP$; $u \in L_p$; $r$ is a prophecy operand, and $q$ has the same lattice as $p$. The last
line lists the prophecy commands.}
\end{figure}

\begin{figure}[tp]
\vspace{1.5em}
\begin{center}
$\infer{\tuple{n, \sigma} \rightarrow n}{}$
\qquad\qquad
$\infer{\tuple{x, \sigma} \rightarrow \sigma(x)}{x \in \dom{\sigma}}$
\qquad\qquad
$\infer{\tuple{x, \sigma} \rightarrow x}{x \notin \dom{\sigma}}$
\\[1.2em]
$\infer{\tuple{\ominus\, a, \sigma} \rightarrow \ominus\, n}{\tuple{a, \sigma} \rightarrow n}$
\qquad\qquad
$\infer{\tuple{\ominus\, a, \sigma} \rightarrow \ominus\, a'}
       {\tuple{a, \sigma} \rightarrow a' \;\;\; a' \notin \Int}$
\\[1.2em]
$\infer{\tuple{a_0 \oplus a_1, \sigma} \rightarrow n_0 \oplus n_1}
       {\tuple{a_0, \sigma} \rightarrow n_0 \;\;\; \tuple{a_1, \sigma} \rightarrow n_1 \;\;\; n_0 \oplus n_1 \mbox{ defined}}
\qquad
\infer{\tuple{a_0 \oplus a_1, \sigma} \rightarrow a_0' \oplus a_1'}
       {\tuple{a_0, \sigma} \rightarrow a_0' \;\;\; \tuple{a_1, \sigma} \rightarrow a_1' \;\;\;
        a_0' \oplus a_1' \notin \Int}$
\end{center}
\vspace{.1in}
\begin{center}
\begin{tabular}{cc}
$\infer{\tuple{l, \sigma, \pi} \rightarrow \tuple{g, \sigma[x \mapsto n], \pi}}
       {P(l) = \cassign{x}{e}{g} \;\;\; \tuple{e, \sigma} \rightarrow n}$
&
$\infer{\tuple{l, \sigma, \pi} \rightarrow \tuple{g, \sigma, \pi}}{P(l) = \cgoto{g}}$
\\[1.2em]
$\infer{\tuple{l, \sigma, \pi} \rightarrow \tuple{g_z, \sigma, \pi}}
       {P(l) = \cifz{x}{g_z}{g_{nz}} \;\;\; \sigma(x) = 0}$
&
$\infer{\tuple{l, \sigma, \pi} \rightarrow \tuple{g_{nz}, \sigma, \pi}}
       {P(l) = \cifz{x}{g_z}{g_{nz}} \;\;\; \sigma(x) \neq 0}$
\end{tabular}
\end{center}
\caption{\label{fig:refsem} Reference Semantics: Expression Evaluation and Command Execution Rules.
$n, n_0, n_1 \in \Int$. In the last expression rule, $a_0' \oplus a_1' \notin \Int$ means that $a_0'$
or $a_1'$ is not an integer or that $\oplus$ is undefined on them.}
\end{figure}

\begin{figure}[tp]

\begin{center}
\begin{tabular}{cc}
\multicolumn{2}{c}{$\infer{\tuple{l, \sigma, \pi} \rightarrow \tuple{g, \sigma, \pi}}
       {P(l) = \ccheck{p}{r}{g} \;\;\; \pi(r) \leq_p \pi(p)}$}
\\[1.2em]
$\infer{\tuple{l, \sigma, \pi} \rightarrow \tuple{g_t, \sigma, \pi}}
       {P(l) = \cbelow{p}{r}{g_t}{g_f} \;\;\; \pi(r) \leq_p \pi(p)}$
&
$\infer{\tuple{l, \sigma, \pi} \rightarrow \tuple{g_f, \sigma, \pi}}
       {P(l) = \cbelow{p}{r}{g_t}{g_f} \;\;\; \pi(r) \not\leq_p \pi(p)}$
\\[1.2em]
\multicolumn{2}{c}{$\infer{\tuple{l, \sigma, \pi} \rightarrow \tuple{g, \sigma[x \mapsto \pi(p)], \pi}}
       {P(l) = \cget{x}{p}{g}}$}
\end{tabular}
\end{center}
\caption{\label{fig:pvsem} Reference Semantics: Prophecy Command Execution Rules. $\pi(r)$ is the
value of the prophecy operand $r$: $\pi(u) = u$ for a value $u$, and $\pi(q)$ for a prophecy variable
$q$.}
\end{figure}

\subsection{Reference Semantics}
\label{sec:refsem}

A store $\sigma : (\VS \cup \VD) \pfun \Int$ maps variables to integers. A \emph{prediction} $\pi$
maps each $p \in \VP$ to a value $\pi(p) \in L_p$ and each value $u$ to itself. The set $\Pi$ of
predictions is ordered pointwise, with least element $\bot$, where $\bot(p) = \bot_p$. A
configuration $\tuple{l, \sigma, \pi}$ has $\dom{\sigma} = \VS \cup \VD$. Expression evaluation
$\tuple{e, \sigma} \rightarrow e'$ and command execution
$\tuple{l, \sigma, \pi} \rightarrow \tuple{l', \sigma', \pi}$ are the smallest relations closed under
the rules of Figures~\ref{fig:refsem} and~\ref{fig:pvsem}. Expression evaluation is total and
deterministic; we write $\sigma(e)$ for its result. Note that evaluation leaves variables outside
$\dom{\sigma}$ in place, so it also serves as a partial evaluator for the first stage.

A configuration $\tuple{l, \sigma, \pi}$ is \emph{final} if $P(l) = \kw{halt}$, \emph{faulting} if
$P(l) = \cassign{x}{e}{g}$ and an operator in $e$ is undefined on the values of its operands in
$\sigma$ (for example, a division by zero), and \emph{mispredicted}, written
$\fails\;\tuple{l, \sigma, \pi}$, if $P(l) = \ccheck{p}{r}{g}$ with $\pi(r) \not\leq_p \pi(p)$. $P$
is \emph{well formed} if $l_0 \in \dom{M}$, $\Succ{P(l)} \subseteq \dom{M}$ for every
$l \in \dom{M}$, $O \subseteq \VD$, and every prophecy command meets the side conditions of
Figure~\ref{fig:corelanguage}. The execution from $\sigma_0$ under $\pi$ starts at $\tuple{l_0, \sigma_0, \pi}$. It
\emph{halts} with \emph{result} $\rho : O \rightarrow \Int$ if it reaches ($\rightarrow^*$) a final
configuration $\tuple{l, \sigma, \pi}$, where $\rho(x) = \sigma(x)$. It \emph{faults} or is
\emph{mispredicted} if it reaches a faulting or a mispredicted configuration. It \emph{diverges} if it
takes $k$ steps ($\rightarrow^k$) for every $k \in \Nat$. In a well-formed program exactly one of
these holds. Two executions \emph{behave the same} if they halt with the same result, both fault, or
both diverge. An \emph{input} $\delta : \VD \rightarrow \Int$ determines the initial store
$\sigma_\delta = \delta \cup \setof{x \mapsto 0 \mid x \in \VS}$.

\begin{definition}[Consistency and meaning]
\label{def:consistent}
A prediction $\pi$ is \emph{consistent} for $\sigma_0$ if the execution from $\sigma_0$ under $\pi$ is
not mispredicted. The \emph{meaning} of $P$ is the set of consistent predictions:
\[
\den{P}(\delta) = \setof{\pi \in \Pi \mid \pi \mbox{ is consistent for } \sigma_\delta}.
\]
\end{definition}

\noindent For $\pi \in \Pi$, $P_\pi$ is $P$ with each $\ccheck{p}{r}{g}$ replaced by $\cgoto{g}$, each
prophecy branch by the $\kw{goto}$ that $\pi$ selects, and each $\cget{x}{p}{g}$ by
$\cassign{x}{\pi(p)}{g}$. If
$\pi$ is consistent for $\sigma_0$, $P$ and $P_\pi$ reach the same configurations from
$\tuple{l_0, \sigma_0, \pi}$.

\subsection{Least Predictions May Not Exist}
\label{sec:noleast}

A natural candidate for the meaning of a prophecy variable is the \emph{least} consistent
prediction. But this least prediction may not exist ---
prophecy branches let predictions determine which checks the execution reaches.
Consider a prophecy variable $p$ over the powerset of $\setof{0,1}$ (implemented as bitmaps) 
and the program $l_0: \cbelow{p}{\setof{0}}{l_2}{l_1},\; l_1:
\ccheck{p}{\setof{1}}{l_2},\; l_2: \kw{halt}$. The prediction $\setof{0}$ skips the check and
$\setof{1}$ passes it, so both are consistent, but $\emptyset$, the only prediction below both, fails
the check. Note that if a well-formed program does not contain prophecy branches or $\kw{get}$ and its
lattices have finite height, a least consistent prediction does exist, and the first stage
computes, when it terminates, the least prediction that satisfies the first stage
(Theorem~\ref{thm:least}; Section~\ref{sec:theorems} defines satisfying the first stage). 
Because \buildit programs use prophecy branches to 
select optimized code (for example, whether to copy a tensor to the GPU in
Section~\ref{sec:einsum}), we take the set $\den{P}(\delta)$ as the meaning and
characterize the prediction the first stage computes in general relative to this set
(Section~\ref{sec:theorems}). 

\subsection{First Stage Execution}
\label{sec:stage1}

The first stage takes a statically independent program $P$ and returns a prediction $\pi_c$ and a
second stage program $Q = \tuple{\emptyset,\; \emptyset,\; \VD \cup \setof{z},\; l_0,\; M,\; O}$ with no
static or prophecy variables, where $z \notin \VS \cup \VD$ is a fresh dynamic variable. It works with \emph{static states} $\tuple{l, \sigma}$, where the static store
$\sigma \in \Sigma_S = \VS \rightarrow \Int$ holds the values of the static variables, and one further
state $\kw{exit}$, the successor of an emitted assignment to $z$ whose evaluation faults, which
the second stage never reaches: $S = (\Lab \times \Sigma_S) \cup \setof{\kw{exit}}$. The initial static state is
$\tuple{l_0, \mathbf{0}}$, where $\mathbf{0}$ maps every static variable to $0$. Commands
$\hat{c} \in \hat{C}$ have static states instead of labels as successors; $\Succ{\hat{c}}$ is defined
as for commands, and for $T : S \pfun \Lab$, $\hat{c}[T]$ is $\hat{c}$ with each successor $s$
replaced by $T(s)$. A \emph{failed prediction} $\fail(p, u) \in \Fail$ records a check on $p$ that
requires the value $u \in L_p$.

\begin{algorithm}[t]
\caption{Driver}
\label{alg:driver}
\begin{algorithmic}[1]
\Procedure{Driver}{} \Comment{returns $\Pi \times \Prog$}
  \State $\var \pi \in \Pi \gets \bot$
  \Loop \Comment{one round: a complete first stage execution under $\pi$} \label{ln:dloop}
    \State $\var T \in S \pfun \Lab \gets \setof{\tuple{l_0, \mathbf{0}} \mapsto l_0}$ \Comment{memo table} \label{ln:ginit}
    \State $\var H \subseteq S \gets \emptyset$ \Comment{memo set} \label{ln:ginitend}
    \State $\var M \in (\Lab \pfun C) \cup \Fail \gets \Call{Execute}{\pi, \epsilon}$ \Comment{run from the start, empty array} \label{ln:gfail}
    \IfThen{$M = \fail(p, u)$}{$\pi \gets \pi[p \mapsto \pi(p) \vee_p u]$} \Comment{raise $p$, reexecute}
    \ElseLine{\Return $(\pi, \tuple{\emptyset, \emptyset, \VD \cup \setof{z}, l_0, M, O})$} \label{ln:gret}
  \EndLoop
\EndProcedure
\end{algorithmic}
\end{algorithm}

\noindent \paragraph{Driver.} Algorithm~\ref{alg:driver} maintains 
the prediction $\pi$, which stores the current values of the prophecy variables.
$\pi$ is initialized to $\bot$ and increases as the first stage encounters 
mispredicted prophecy variable checks. The driver executes \emph{rounds}, with each round invoking
Algorithm~\ref{alg:execute}, which executes the first stage until it either completes or encounters
a prophecy variable misprediction, which then triggers a prophecy variable update and new round
with a new first stage reexecution. 

\begin{algorithm}[t]
\caption{Execute}
\label{alg:execute}
\begin{algorithmic}[1]
\Procedure{Execute}{$\pi \in \Pi,\; D \in \mathbb{B}^*$} \Comment{updates $T$, $H$; returns $(\Lab \pfun C) \cup \Fail$}
  \State $\var s \in S \gets \tuple{l_0, \mathbf{0}}$ \Comment{current static state}
  \State $\var R \subseteq S \gets \emptyset$ \Comment{static states executed so far}
  \State $\var N \subseteq S \gets \emptyset$ \Comment{static states executed after the replay}
  \State $\var k \in \Nat \gets 0$ \Comment{dynamic branches taken so far}
  \State $\var F \in \Lab \pfun C \gets \emptyset$ \Comment{the program this execution generates}
  \Loop
    \IfThen{$s \in R$}{\textbf{break}} \Comment{loop back} \label{ln:eloop}
    \IfThen{$k = |D|$ and $s \in H$}{\textbf{break}} \Comment{memoization hit} \label{ln:ememo}
    \State $\var \hat{c} \in \hat{C} \cup \Fail \gets \Call{Expand}{\pi, s}$ \label{ln:gexpand}
    \IfThen{$\hat{c} = \fail(p, u)$}{\Return $\fail(p, u)$} \label{ln:efail}
    \State $R \gets R \cup \setof{s}$
    \If{$k = |D|$} \Comment{after the replay: generate the code of $s$}
      \State $N \gets N \cup \setof{s}$ \label{ln:gnew}
      \ForAll{$s' \in \Succ{\hat{c}}$ with $s' \notin \dom{T}$} \label{ln:gsucc}
        \State $T(s') \gets l_{|T|}$ \Comment{fresh label}
      \EndFor \label{ln:gsuccend}
      \State $F(T(s)) \gets \hat{c}[T]$ \Comment{successors replaced by their labels} \label{ln:gstore}
    \EndIf
    \If{$\hat{c} = \kw{halt}$} \textbf{break}
    \ElsIf{$\hat{c} = \cifz{x}{s_z}{s_{nz}}$} \Comment{a dynamic branch} \label{ln:edyn}
      \If{$k = |D|$} \Comment{a new branch: explore both arms from the start} \label{ln:enew}
        \State $F_{\kw{true}} \gets \Call{Execute}{\pi, D \cdot \kw{true}}$ \label{ln:erec}
        \IfThen{$F_{\kw{true}} = \fail(p, u)$}{\Return $\fail(p, u)$}
        \State $F_{\kw{false}} \gets \Call{Execute}{\pi, D \cdot \kw{false}}$
        \IfThen{$F_{\kw{false}} = \fail(p, u)$}{\Return $\fail(p, u)$} \label{ln:erecfalse}
        \State $F \gets F \cup F_{\kw{true}} \cup F_{\kw{false}}$ \Comment{merge the programs of both arms} \label{ln:erecend}
        \State \textbf{break}
      \EndIf
      \State $s \gets \kw{if}\; D[k]\; \kw{then}\; s_{nz}\; \kw{else}\; s_z$;\; $k \gets k + 1$ \Comment{$\kw{true}$ selects the nonzero arm} \label{ln:edynend}
    \Else\ $s \gets s'$, where $\Succ{\hat{c}} = \setof{s'}$
    \EndIf
  \EndLoop
  \State $H \gets H \cup N$ \Comment{update the memo set before returning} \label{ln:ememoupd}
  \State \Return $F$ \label{ln:eret}
\EndProcedure
\end{algorithmic}
\end{algorithm}

\noindent \paragraph{Execute.}
Following the \buildit dynamic control flow exploration algorithm~\cite{buildit,ajay-thesis}, 
Algorithm~\ref{alg:execute} uses repeated program execution to explore
both sides of dynamic branches. Each execution starts with a bitvector 
$D$ that specifies directions to follow for dynamic branches encountered in previous executions 
(line~\ref{ln:edynend}). If the execution exhausts the bitvector and encounters a new dynamic branch, it
recursively reexecutes the program, once for each direction of the newly encountered dynamic branch
(lines~\ref{ln:erec}--\ref{ln:erecfalse}). Each of these executions returns the program for its arm of the newly
encountered dynamic branch. The algorithm merges the 
two generated programs, prepends the program explored before the newly encountered dynamic branch, 
and returns the generated second stage program (lines~\ref{ln:erecend} and~\ref{ln:eret}). 
Algorithm~\ref{alg:execute} works with the following data structures:
\begin{itemize}
\item \textbf{Current Static State} $s$: it starts at $\tuple{l_0, \mathbf{0}}$. Algorithm~\ref{alg:execute}
generates code for each static state it executes after its replay (line~\ref{ln:gstore}). 
\item \textbf{Direction Array} $D \in \mathbb{B}^*$: the execution's path. The $k$-th dynamic branch
  it reaches takes direction $D[k]$ ($\kw{true}$ selects the nonzero arm). $k$ counts the
  dynamic branches taken so far. 
\item \textbf{Memo Table} $T : S \pfun \Lab$: the label of the code for each static state, in the
  second stage program. All executions of a round share $T$; the driver initializes it at the start
  of each round to map only $\tuple{l_0, \mathbf{0}}$ to $l_0$ (line~\ref{ln:ginit}).
\item \textbf{Memo Set} $H \subseteq S$: the static states that finished executions executed after
  their replays, whose code has therefore already been generated. An execution stops when it
  reaches one of the static states in $H$ after the replay. All executions of a round share $H$; the driver initializes
  it to $\emptyset$ at the start of each round (line~\ref{ln:ginitend}).
\item \textbf{Executed States} $R \subseteq S$: the static states the execution has executed, and
  $N \subseteq R$, those it executed after its replay.
\item \textbf{Program} $F : \Lab \pfun C$: the code the execution generates.
\end{itemize}

Prophecy variables extend the standard BuildIt algorithm described above by 
passing failures caused by mispredicted prophecy variable checks
back to Algorithm~\ref{alg:driver}, which updates the prediction $\pi$ and reinvokes
Algorithm~\ref{alg:execute} with the new prediction $\pi$, with only $\pi$ retained across rounds.
Each round, the driver invokes \textsc{Execute} on the empty array, which returns either the code
$M$ of the second stage program or a failed prediction $\fail(p, u)$ (line~\ref{ln:gfail}). On a
failed prediction the driver raises $p$ to $\pi(p) \vee_p u$ and starts a new round, which
reexecutes $P$ with the raised prediction. Otherwise it returns the successful prediction $\pi$ and
the second stage program (line~\ref{ln:gret}).

While $k < |D|$, an execution \emph{replays} the path of its parent and generates nothing
(the code for this path has already been generated). Once
$k = |D|$ it executes static states that are in neither $R$ nor $H$. 
For each static state $s$ it executes, it adds $s$ to $N$, gives each
new successor of $\hat{c}$ a fresh label in $T$, and adds $\hat{c}[T]$ to $F$ at $T(s)$
(lines~\ref{ln:gnew}--\ref{ln:gstore}). It stops at $\kw{halt}$, at a failed check
(line~\ref{ln:efail}), or in one of three other ways:
\begin{itemize}
\item \textbf{Loop Back:} $s \in R$ (line~\ref{ln:eloop}). The code of $s$ links back to its label,
  closing the loop.
\item \textbf{Memoization Hit:} $k = |D|$ and $s \in H$ (line~\ref{ln:ememo}). $H$ holds the static
  states that finished executions executed after their replays, where an execution finishes only
  after all of the executions it started. The code from $s$ on therefore already exists, and the
  label $T(s)$ links to it.
\item \textbf{New Dynamic Branch:} $\hat{c}$ is a dynamic branch and $k = |D|$ (line~\ref{ln:enew}).
  The execution invokes \textsc{Execute}, depth first, on $D \cdot \kw{true}$ and $D \cdot \kw{false}$
  (lines~\ref{ln:erec}--\ref{ln:erecfalse}). Each of these executions replays its parent up to the
  branch, takes its own arm, and returns the program for that arm. The execution merges the two
  programs into $F$ (line~\ref{ln:erecend}).
\end{itemize}
Just before it returns $F$, an execution adds $N$ to $H$ (line~\ref{ln:ememoupd}). In effect, the
program for $D$ is the code along $D$'s new path, the branch at its end, and the merged programs of
the two arms. \textsc{Execute} on the empty array therefore returns the whole second stage program.

\begin{algorithm}[t]
\caption{Expand}
\label{alg:expand}
\begin{algorithmic}[1]
\Procedure{Expand}{$\pi \in \Pi,\; s \in S$} \Comment{returns $\hat{C} \cup \Fail$}
  \IfThen{$s = \kw{exit}$}{\Return $\kw{halt}$}
  \State $\var \tuple{l, \sigma} \in \Lab \times \Sigma_S \gets s$
  \If{$P(l) = \cassign{x}{e}{g}$ and $x \in \VS$} \label{ln:sassign}
    \IfThen{$\sigma(e) = n \in \Int$}{\Return $\cgoto{\tuple{g, \sigma[x \mapsto n]}}$}
    \ElseLine{\Return $\cassign{z}{\sigma(e)}{\kw{exit}}$} \Comment{faults in the second stage} \label{ln:sassignfault}
  \ElsIfLine{$P(l) = \cassign{x}{e}{g}$ and $x \in \VD$}{\Return $\cassign{x}{\sigma(e)}{\tuple{g, \sigma}}$} \label{ln:dassign} \label{ln:dassignret}
  \ElsIf{$P(l) = \cifz{x}{g_z}{g_{nz}}$ and $x \in \VS$} \label{ln:sifz}
    \IfThen{$\sigma(x) = 0$}{\Return $\cgoto{\tuple{g_z, \sigma}}$}
    \ElseLine{\Return $\cgoto{\tuple{g_{nz}, \sigma}}$} \label{ln:sifzend}
  \ElsIf{$P(l) = \cifz{x}{g_z}{g_{nz}}$ and $x \in \VD$} \label{ln:difz}
    \State \Return $\cifz{x}{\tuple{g_z, \sigma}}{\tuple{g_{nz}, \sigma}}$ \label{ln:difzret}
  \ElsIfLine{$P(l) = \cgoto{g}$}{\Return $\cgoto{\tuple{g, \sigma}}$}
  \ElsIfLine{$P(l) = \kw{halt}$}{\Return $\kw{halt}$}
  \ElsIfLine{$P(l) = \cget{x}{p}{g}$}{\Return $\textsc{Expand}$ of $\cassign{x}{\pi(p)}{g}$ at $\tuple{l, \sigma}$} \label{ln:get}
  \ElsIf{$P(l) = \ccheck{p}{r}{g}$}
    \IfThen{$\pi(r) \leq_p \pi(p)$}{\Return $\cgoto{\tuple{g, \sigma}}$} \label{ln:checkok}
    \ElseLine{\Return $\fail(p, \pi(r))$} \label{ln:checkfail}
  \ElsIf{$P(l) = \cbelow{p}{r}{g_t}{g_f}$} \label{ln:pbranch}
    \IfThen{$\pi(r) \leq_p \pi(p)$}{\Return $\cgoto{\tuple{g_t, \sigma}}$}
    \ElseLine{\Return $\cgoto{\langle g_f, \sigma\rangle}$} \label{ln:pbranchend}
  \EndIf
\EndProcedure
\end{algorithmic}
\end{algorithm}

\paragraph{Expand.} Algorithm~\ref{alg:expand} works on a single static state $\tuple{l, \sigma}$:
the label $l$ of the command to execute and the static store $\sigma$. Under the prediction $\pi$ it
returns either a command $\hat{c}$ whose successors are static states or a failed prediction. The
basic idea is that the first stage does not have the dynamic values, so it executes the code that
depends only on $\sigma$ and $\pi$ and emits the rest:
\begin{itemize}
\item \textbf{Static Code Executes:} An assignment to a static variable updates $\sigma$ and becomes
  a $\kw{goto}$ (lines~\ref{ln:sassign}--\ref{ln:sassignfault}). If it faults, it is emitted as
  an assignment to $z$ with successor $\kw{exit}$, so that the second stage faults at the same point
  without mentioning the static variable. A branch on a static
  variable becomes a $\kw{goto}$ to the arm $\sigma$ selects (lines~\ref{ln:sifz}--\ref{ln:sifzend}).
\item \textbf{Dynamic Code Is Emitted:} An assignment to a dynamic variable is emitted with its
  expression partially evaluated to $\sigma(e)$ (line~\ref{ln:dassign}). A
  branch on a dynamic variable is emitted with both arms (lines~\ref{ln:difz}--\ref{ln:difzret}).
\item \textbf{Prophecy Commands Execute Under $\pi$:} A check that $\pi$ satisfies becomes a
  $\kw{goto}$ (line~\ref{ln:checkok}); a check that $\pi$ fails returns $\fail(p, \pi(r))$
  (line~\ref{ln:checkfail}). A prophecy branch becomes a $\kw{goto}$ to the arm that $\pi$ selects
  (lines~\ref{ln:pbranch}--\ref{ln:pbranchend}). A $\cget{x}{p}{g}$ executes as
  $\cassign{x}{\pi(p)}{g}$ (line~\ref{ln:get}).
\end{itemize}

\subsection{Theorems}
\label{sec:theorems}

We have mechanized the formalism and the theorems in Lean~4. Supervised by one of the authors, the coding agent Claude Code 
wrote all of the code and proofs in the mechanization. 
The mechanization builds on a verified base compiler (about 1,450 lines of
definitions, 4,600 lines of proofs, and 405 theorems) that generates AArch64 assembler. A verified
extension (about 1,500 lines of definitions, 6,200 lines of proofs, and 425 theorems) adds static,
dynamic, and prophecy variables, with the verification extending down to the AArch64 machine model.
(We count non-blank, non-comment lines.) The trusted base
consists of the parsers, the assembly printer, %
the AArch64 machine model, the
operational semantics, the compiler drivers, and infrastructure such as Lean~4, the assembler, the
linker, the C runtime, and the operating system. The project contains no \texttt{sorry} or \texttt{admit}, and every theorem
depends only on the three standard Lean axioms (\texttt{propext}, \texttt{Classical.choice}, and
\texttt{Quot.sound}).

Let $P$ be well formed and statically independent. Theorem~\ref{thm:termination} gives conditions
under which Algorithm~\ref{alg:driver} terminates and returns. The remaining theorems assume that it 
terminates and returns
$(\pi_c, Q)$. Below each theorem we name the corresponding Lean theorem and the file in which it appears; file
names are relative to the \texttt{ProphecyVariableReexecute/} directory of the artifact.

\begin{theorem}[Termination]
\label{thm:termination}
If every lattice $L_p$ has finite height at most $h_p$ and each round of Algorithm~\ref{alg:driver}
terminates under every prediction, Algorithm~\ref{alg:driver} terminates after at most
$1 + \sum_{p \in \VP} h_p$ rounds.
\end{theorem}

\leanname{\texttt{first\_stage\_terminates\_height} and \texttt{first\_stage\_runs\_le\_height}}{Staged/Termination.lean}

\noindent The \emph{height} $h(L)$ of a lattice $L$ is the largest $n$ for which $L$ contains a chain
$a_0 < a_1 < \cdots < a_n$. Each failed round strictly raises one prophecy variable, because the
failed check requires a value that the current value does not cover. The successive values of each
variable $p$ therefore form a strictly increasing chain from $\bot_p$, so the driver raises $p$ at
most $h(L_p) \leq h_p$ times. Note that the bound requires only finite height, not finite lattices. 
Like any partial evaluator, a round may fail to
terminate, for example on a static loop that never terminates or on a dynamic loop whose
static state never repeats. 

\begin{theorem}[Least upper bound of the raises]
\label{thm:lub}
For every $p \in \VP$, $\pi_c(p)$ is the least element of $L_p$ that is above every value $u$ for which
Algorithm~\ref{alg:driver} raised $p$, and each such $u$ is the value $\pi(r)$ that a check
$\ccheck{p}{r}{g}$ required in a round, under prediction $\pi$, in which the check failed.
\end{theorem}

\leanname{\texttt{first\_stage\_lub}}{Staged/Prophecy.lean}

\noindent In particular, each raise is the least value above both the variable's current value and the
value that the failed check required.

\begin{theorem}[Consistency]
\label{thm:consistency}
For every input $\delta$, $\pi_c \in \den{P}(\delta)$.
\end{theorem}

\leanname{\texttt{first\_stage\_correct}, first conjunct}{Staged/Prophecy.lean}

\noindent The first stage explores both arms of every dynamic branch. 
Every execution of $P$ on every input follows one of the resulting explored paths. 
The final round fails no check on any of these paths, so no execution of $P$ under $\pi_c$ fails a check.

\begin{theorem}[Second stage correctness]
\label{thm:stage2}
For every input $\delta$, the second stage program $Q$ and the program $P_{\pi_c}$, both executed under
the reference semantics from $\sigma_\delta$, behave the same: $Q$ halts with result $\rho$ if and only
if $P_{\pi_c}$ halts with result $\rho$, $Q$ faults if and only if $P_{\pi_c}$ faults, and $Q$ diverges
if and only if $P_{\pi_c}$ diverges.
\end{theorem}

\leanname{\texttt{first\_stage\_correct}, second conjunct}{Staged/Prophecy.lean}

\noindent Because $\pi_c$ is consistent (Theorem~\ref{thm:consistency}), $P_{\pi_c}$ and $P$ under
$\pi_c$ reach the same configurations, $Q$ behaves exactly as $P$ under $\pi_c$. Conceptually, the
proof is a lockstep forward simulation that uses the memo table $T$ of the final round. It relates a
configuration $\tuple{l, \sigma, \pi_c}$ of $P_{\pi_c}$ to a configuration $\tuple{k, \sigma_Q, \pi}$
of $Q$ when $T(\tuple{l, \sigma'}) = k$, where $\sigma'$ is the restriction of $\sigma$ to $\VS$, and
$\sigma_Q$ agrees with $\sigma$ on $\VD$. $Q$ has no prophecy variables, so $\pi$ is irrelevant. Each
step of $P_{\pi_c}$ is matched by exactly one step of $Q$. The reverse direction follows because the
three outcomes are exhaustive for $P_{\pi_c}$ and mutually exclusive for $Q$.

\begin{theorem}[Static variables are eliminated]
\label{thm:nostatic}
No command of $Q$ reads or assigns a static variable.
\end{theorem}

\leanname{\texttt{first\_stage\_no\_static}}{Staged/Resolve.lean}

\noindent \textsc{Expand} emits a branch only on a dynamic variable, partially evaluates every emitted
expression under $\sigma$, which replaces each static variable with its value, and emits assignments
only to dynamic variables and $z$.

Without prophecy branches and $\kw{get}$, the first stage computes a least prediction in a precise
sense. We say that a static state $s$ is \emph{reachable} under a prediction $\pi$ if $s = \tuple{l_0, \mathbf{0}}$ or
$s \in \Succ{\textsc{Expand}(\pi, s')}$ for some static state $s'$ reachable under $\pi$ (with
$\Succ{\fail(p, u)} = \emptyset$), and that
$\pi$ \emph{satisfies the first stage} if $\textsc{Expand}(\pi, s) \notin \Fail$ for every static
state $s$ reachable under $\pi$.

\begin{theorem}[Least prediction for the first stage]
\label{thm:least}
$\pi_c$ satisfies the first stage. If $P$ has no prophecy branches and no $\kw{get}$, then
$\pi_c \leq \pi$ for every prediction $\pi$ that satisfies the first stage.
\end{theorem}

\leanname{\texttt{first\_stage\_satisfies} and \texttt{first\_stage\_least}}{Staged/Least.lean}

\noindent All prophecy variable checks pass in the final first stage round, with the memo table
for this round containing every static state reachable under $\pi_c$. $\pi_c$ 
therefore satisfies the first stage. Without prophecy branches and $\kw{get}$, a
prediction changes the successors of a static state only by failing a check at that static state. 
A $\kw{get}$ would place the prediction in the static store, and from there into the control flow. 
Every static state reachable under any prediction is therefore reachable under each $\pi$ that satisfies the first
stage, so $\pi$ satisfies the check at which each failed round stops and each raise stays below
$\pi$. Because the driver starts at $\bot$, $\pi_c \leq \pi$.

Finally, composing Theorem~\ref{thm:stage2} with our verified back end (normalization, cleanup, and
code generation to an AArch64 machine model) extends the guarantee to the compiled machine code. If
the observable variables of $P$ are program variables rather than compiler temporaries, then for every
input that is zero on the variables the compiled program does not mention, the machine code halts
with the same observable values as $P_{\pi_c}$, faults if and only if $P_{\pi_c}$ faults, and
diverges if and only if $P_{\pi_c}$ diverges (\texttt{first\_stage\_asm\_correct} in
\texttt{Staged/ProphecyLang.lean}).

    \section{Applications}
\label{sec:applications}

We next present the use of BuildIt prophecy and history 
variables for optimizing computations written in a BuildIt 
DSL for performing einsum tensor computation and
a BuildIt DSL for constructing neural networks. We also present the use of prophecy variables to implement a binding time analysis to identify static and dynamic variables in BuildIt programs.
We present an overview of BuildIt (Section~\ref{sec:buildit}), 
prophecy variables in BuildIt (Section~\ref{sec:builditprophecy})
and then the two DSLs along with a benchmark application
in each DSL (Sections~\ref{sec:einsum} and \ref{sec:nn}).
Section~\ref{sec:binding} presents the binding time analysis and results.
We include performance results for each of the DSL computations. 
The BuildIt implementation is
publicly available (repository link omitted for double-blind review). 

\subsection{\buildit Overview}
\label{sec:buildit}
\buildit is a lightweight DSL implementation system~\cite{buildit, ajay-thesis} that uses multi-stage execution to generate optimized programs. \buildit takes a purely library approach and does not require any special language or runtime support. As a result, the first stage executes as standard C++ code that generates optimized C, C++, and/or CUDA second stage code. This generated second stage code can then be independently compiled and executed multiple times with no need to reexecute the first stage. In a typical \buildit program the vast majority of the computation is performed by the second stage. 

The \buildit library introduces two new type-templates - \dyn{T} and \static{T}. Variables and expressions declared as \static{T} are evaluated entirely in the first stage including all side effects and control flow constructs like if-then-else and while loops. Variables and expressions declared as \dyn{T} are generated as it is to be executed in the second stage. Figure~\ref{lst:powerex} a) shows an example \buildit program with the \static{T} and \dyn{T} types. The power function shown has the base declared as \dyn{int} and hence all operations on it are generated to be executed in the second stage. The exponent is declared as \static{int} and is supplied as \texttt{15} on Line 12. Figure~\ref{lst:powerex} b) shows the corresponding generated second stage code. Note that all operations on the exponent including the condition on Line 4 and the loop on Line 3 are completely evaluated to only preserve an unrolled loop with operations on the \dyn{int} variables.

\subsubsection{Reexecution in \buildit}
Since \buildit takes a purely library-based approach, it does not have access to a program AST or the source text. \buildit extracts operations on \dyn{T} variables by overloading all operators including assignment and all binary operations. \static{T} is a wrapper around \texttt{T} with some extra book-keeping logic. However, overloading control-flow constructs like if-then-else and while loops is not possible in C++, so \buildit uses a reexecution based mechanism where the program to be partially evaluated is executed multiple times to explore different dynamic branches (if-then-else and loops where the condition depends on at least one \dyn{T} variable). \buildit includes optimizations like memoization and static tags based indexing to avoid exponential blowup in the extraction process~\cite{buildit}. This existing reexecution machinery allows us to seamlessly implement Algorithm~\ref{alg:driver} without adding any significant complexity. 

Prior work has already shown how the \static{T} and \dyn{T} types can be used to propagate information about the program in the forward direction~\cite{buildit,ajay-thesis}. \static{T} variables essentially act like history variables that can pass information between different functions implemented as part of the DSL~\cite{buildsl}. These DSLs implement forward data-flow analyses using these history variables and enable transformations that improve performance. However, it is not possible to implement traditional backward analyses that capture information about the future execution and propagate it backwards against the flow of control using just \static{T} and \dyn{T}. This is a major limitation of the \buildit system and restricts analyses like liveness, fusion, and opportunistic data movement among others. 

\begin{figure}
\begin{subfigure}[b]{0.6\textwidth}
\begin{lstlisting}[language=buildit]
dyn_var<int> power (dyn_var<int> base, static_var<int> exp) {
  dyn_var<int> res = 1, x = base;
  while (exp > 1) {
    if (exp % 2 == 1)
      res = res * x;
    x = x * x;
    exp = exp / 2;
  }
  return res * x;
}    
...
auto program = ctx.extract_function(power, "power_15", 15);
\end{lstlisting}
\end{subfigure}
\hspace{2em}
\begin{subfigure}{0.3\textwidth}
\begin{lstlisting}[language=buildit]
int power_15 (int arg0) {
  int var0 = 1;
  int var1 = arg0;
  var0 = var0 * var1;
  var1 = var1 * var1;
  var0 = var0 * var1;
  var1 = var1 * var1;
  var0 = var0 * var1;
  var1 = var1 * var1;
  int var2 = var0 * var1;
  return var2;
}
\end{lstlisting}
\end{subfigure}
    \caption{a) Implementation of a power function that returns $base^{exp}$ using repeated squaring. The base is declared as \dyn{int} and thus will be input in the second stage. The exponent is declared as \static{int} and is provided in the driver call as 15. The generated program will be specialized for this value of exponent. b) Generated second stage program for the input program in a). Notice this program doesn't have any control-flow based on the exponent that has been evaluated completely in the first stage.}
    \label{lst:powerex}
\end{figure}

\begin{figure}
\begin{lstlisting}[language=buildit]
template <typename T>
class prophecy_var {
    // Private members and functions
    ...
public:
    template <typename...Args>
    prophecy_var(Args&&...args) {
        ...
    }  
    void assert_requires_atleast (T::value_type e) {
        ...
    }
    bool is_at_most(T::value_type e) {
        ...
    }
    T::value_type get (void) {
        ...
    }
}; 
\end{lstlisting}
\caption{\label{fig:api} BuildIt prophecy variable API including constructor and lattice operation interface.}
\end{figure}

\subsection{Prophecy Variables in BuildIt}
\label{sec:builditprophecy}
We extend \buildit to introduce a new type template, \bldit{prophecy_var<T>} (see BuildIt prophecy variable API in Figure~\ref{fig:api}). Like \bldit{static_var<T>} variables, \bldit{prophecy_var<T>} variables are completely evaluated in the first stage with the exception that the values of updated \bldit{prophecy_var<T>} variables persist across reexecutions. 
The \bldit{prophecy_var<T>} interface requires: 1) the wrapped type \texttt{T} is a lattice and 2) all prophecy variable updates implement lattice joins. \bldit{static_var<T>} variables can be used to record and propagate first stage information about the past program execution to program transformation and optimization sites; \bldit{prophecy_var<T>} variables can be used at program transformation and optimization sites to predict information about the future program execution. Both history and prophecy variables can be used to conditionally generate different code. Together, \bldit{static_var<T>} and \bldit{prophecy_var<T>} variables can be used to collect information
about both the previous and future execution of the program, then transfer that information to relevant
transformation and optimization sites, eliminating the 
need for traditional program analysis algorithms
that operate on intermediate program representations. 

Our implementation of \prophecy{T} leverages the existing reexecution machinery while preserving prophecy var updates across runs. This implementation has been formalized in Section~\ref{sec:formalization}.

\subsection{Tensor Data Movement for GPU Kernels}
\label{sec:einsum}
In this section we describe an application of prophecy variables that fixes a critical shortcoming in \buildit allowing performance critical optimizations that traditionally required backwards analysis on an explicit IR. 
This benchmark evaluates a \buildit DSL for arbitrary multidimensional tensor operations expressed as einsum expressions~\cite{einsum-buildit}. Tensor shapes and sizes are first-stage values. Implemented as a C++ library using \buildit types, its first stage runs in standard C++ to generate second-stage tensor code, including CUDA GPU kernels. The key type, \texttt{tensor<E>}, uses the first-stage template parameter \texttt{E} for the element type, stores dimensions as \static{std::vector<int>}, and second-stage values as \dyn{E*}. Operators \texttt{[]}, \texttt{+}, \texttt{*}, \texttt{=}, \texttt{+=}, and \texttt{*=} implement einsum operations.

Figure~\ref{lst:einsum} shows matrix multiplication in the DSL: \bldit{run_on_gpu(...)} marks \bldit{C[i][j]} \bldit{+= A[i][k]} \bldit{* B[k][j]} for GPU kernel generation. The complete DSL is in supplementary Section~\ref{sec:appeinsum}; simplified implementations appear in Figures~\ref{lst:einsumconstructor}, \ref{lst:einsumrunongpu}, and \ref{lst:einsumassign}. The implementation uses only \static{T}, \dyn{T}, and \prophecy{T}, without requiring a parser, IR, or custom code generation.

GPU buffers must be allocated and CUDA copies inserted before kernel launch, although which tensors will be accessed or read is known only after encountering the kernel body. Prophecy variables resolve this dependency. Figure~\ref{lst:einsumconstructor} shows the tensor class and constructor, with two \texttt{TrueTop} prophecy variables (as in Figure~\ref{fig:filehandlepredict}): \texttt{needs\_gpu} predicts whether a tensor is ever needed on the GPU, and \texttt{gpu\_read} whether it is read by a particular kernel. The latter is allocated later through a pointer because it is call-site-specific. The class holds host and GPU buffers; on Lines~8--14, the constructor conditionally emits \texttt{cuda\_malloc} when \texttt{needs\_gpu} is true.

Figure~\ref{lst:einsumrunongpu} shows \texttt{run\_on\_gpu}, used in Figure~\ref{lst:einsum} to wrap GPU code. For each live tensor, it creates a prophecy variable of type \prophecy{TrueTop} to predict reads in the next kernel, copying only those predicted to be read. It then sets the global \static{enum device type} variable \texttt{current\_device} to \texttt{GPU}, executes the body, restores \texttt{HOST}, and deletes the prophecy variables.

Figure~\ref{lst:einsumassign} shows \texttt{operator=} and \texttt{get\_value} for tensor writes and reads. On the GPU, both assert \texttt{needs\_gpu} is true; reads also assert \texttt{gpu\_read} is true, completing the prophecy-variable predictions and assertions.

\begin{figure}
\small
\begin{lstlisting}[language=buildit]
void matrix_mul(int M, int N, int O) {
    tensor<float> A({M, N});
    tensor<float> B({N, O});
    tensor<float> C({M, O});
    index i, j, k;
    A[i][j] = 3.0;
    B[i][j] = 4.0 + i + j;
    run_on_gpu([&]() {
        C[i][j] += A[i][k] * B[k][j];
    });
}
\end{lstlisting}
\caption{Matrix-matrix multiply in the einsum DSL. The two inputs and the outputs are of sizes \texttt{M x N}, \texttt{N x O} and \texttt{M x O} respectively where \texttt{M, N} and \texttt{O} are first stage parameters.}
\label{lst:einsum}
\end{figure}

\begin{figure}
\begin{lstlisting}[language=buildit]
template <typename T>
class tensor {
    // Two prophecy variables for predicting allocation and copying decisions
    prophecy_var<TrueTop> needs_gpu;
    prophecy_var<TrueTop> *gpu_read;
    // Buffers for host and GPU
    dyn_var<float*> host_buffer, gpu_buffer;
    tensor(dyn_var<float*> h_buffer, std::vector<size_t> sizes): needs_gpu(TrueTop::FALSE), 
                host_buffer(h_buffer) { 
        // Conditionally allocate gpu buffer if the tensor is ever referenced on the GPU
        if (!needs_gpu.is_at_most(TrueTop::FALSE)) {
            gpu_buffer = runtime::cuda_malloc(sizeof(T) * total_elems(sizes));
        } 
    } 
};
\end{lstlisting}
\caption{\label{lst:einsumconstructor} Implementation of the tensor class and the constructor.}
\vspace{1em}
\begin{lstlisting}[language=buildit]
void run_on_gpu(std::function<void(void)> body) {
    // For each tensor define a new prophecy var to check if it is read in the next kernel
    // If yes, copy the tensor
    for (auto &t: registered_tensors) {
        t.gpu_read = new prophecy_var<TrueTop>(TrueTop::FALSE);
        if (!t.gpu_read->is_at_most(TrueTop::FALSE)) {
            runtime::cuda_memcpy(t.gpu_buffer, t.host_buffer, t.total_size);
        }
    }
    current_device = GPU;
    body();
    current_device = HOST;
    // Delete the prophecy variables
    for (auto &t: registered_tensors) {
        delete t.gpu_read;
    } 
}
\end{lstlisting}
\caption{\label{lst:einsumrunongpu} Implementation of the \texttt{run\_on\_gpu} operator for marking operations to be generated for executing on the GPU.}
\vspace{1em}
\begin{lstlisting}[language=buildit]
template <typename T>
void tensor_access<T>::operator=(tensor_term<T> &rhs) {
    // If running on GPU, mark the tensor to be required on GPU but not read
    if (current_device == GPU) {
        m_tensor.needs_gpu.assert_requires_atleast(TrueTop::TRUE);
    }
    ...
}
template <typename T>
dyn_var<T> tensor_term<T>::get_value(void) {
    // If running on GPU, mark the tensor to be read on GPU and required on GPU
    if (current_device == GPU) {
        m_tensor.needs_gpu.assert_requires_atleast(TrueTop::TRUE);
        m_tensor.gpu_read->assert_requires_atleast(TrueTop::TRUE);
        return m_tensor.gpu_buffer[m_index];
    }
    return m_tensor.host_buffer[m_index];
}
\end{lstlisting}
\caption{\label{lst:einsumassign} Implementation of the tensor operators for assignment and \texttt{[]}.}
\label{lst:einsummini}
\end{figure}

\begin{table}
\begin{minipage}[t]{0.45\textwidth}
\centering
\small
\begin{tabular}{|l|r|r|r|}
\hline
Size & Copy & Unified & Prophecy \\  
     & All  & Memory  &  Enabled \\ \hline
\hline
\multicolumn{4}{|c|}{First Stage}\\ \hline
All sizes & 9,212 & 8,336 & 38,335 \\ \hline \hline
\multicolumn{4}{|c|}{Second Stage}\\ \hline

64x64           & 149 & 239 & \textbf{43} \\ \hline
128x128         & 284 & 269 & \textbf{94} \\ \hline
256x256         & 840 & 615 & \textbf{248} \\ \hline
512x512         & 3053& 1647& \textbf{1059} \\ \hline
1024x1024       & 9,863 & 6,343 & \textbf{4,244} \\ \hline
2048x2048       & 46,552 & 32,395 & \textbf{26,803} \\ \hline
4096x4096       & 380,378 & 316,443 & \textbf{307,830} \\
\hline
\end{tabular}
\caption{\label{tbl:matrixmatrix} Execution time in microseconds for matrix-matrix multiplication with the three strategies - copying all tensors to GPU memory, using unified memory, and BuildIt with prophecy variables, which copies only accessed tensors to GPU memory. We report minimum execution times in microseconds across 10 runs. Typical applications compile the kernel once (this compilation includes the first stage execution), then run hundreds of iterations of the compiled kernel.}
\end{minipage}
$\;\;\;$
\begin{minipage}[t]{0.45\textwidth}
\centering
\small
\begin{tabular}{|l|r|r|r|}
\hline
Size & Copy & Unified & Prophecy \\  
     &    All      & Memory  &  Enabled \\ \hline
\hline
\multicolumn{4}{|c|}{First Stage}\\ \hline
All sizes & 7,154 & 6,180 & 27,167 \\ \hline \hline
\multicolumn{4}{|c|}{Second Stage}\\ \hline
64x64           & 112 & 79 & \textbf{35} \\ \hline
128x128         & 173 & 132 & \textbf{60} \\ \hline
256x256         & 362 & 237 & \textbf{129} \\ \hline
512x512         & 1,199 & 621 & \textbf{468} \\ \hline
1024x1024       & 3,061 & 1,356 & \textbf{1,054} \\ \hline
2048x2048       & 9,647 & 3,564 & \textbf{2,854} \\ \hline
4096x4096       & 34,341 & 9,754 & \textbf{9,199} \\ \hline
\end{tabular}
\caption{\label{tbl:matrixvector} Execution time in microseconds for matrix-vector multiplication with the three strategies - copying all tensors to GPU memory, using unified memory, and BuildIt with prophecy variables, which copies only accessed tensors to GPU memory. We report minimum execution times in microseconds across 10 runs. Typical applications compile the kernel once (this compilation includes the first stage execution), then run hundreds of iterations of the compiled second stage kernel with no additional first stage executions.
}
\end{minipage}

\vspace{.1in}
\centering
\small
\begin{tabular}{|l|r|r|r|}
\hline
{Matrix-Matrix Multiply} & Copy All & Unified Memory & Prophecy Variable \\ \hline 
Prophecy Variable Reexecutions & - & - & 7 \\ \hline
Dynamic Control Flow Executions & 25 & 25 & 113 \\ \hline
First Stage Execution Time & 9,212 & 8,336 & 38,335 \\ \hline \noalign{\vskip 8pt} \hline
{Matrix-Vector Multiply} & Copy All & Unified Memory & Prophecy Variable \\ \hline 
Prophecy Variable Reexecutions & - & - & 7 \\ \hline
Dynamic Control Flow Executions & 19 & 19 &  89 \\ \hline
First Stage Execution Time &  7,154 & 6,180 & 27,167 \\ \hline 
\end{tabular}
\caption{Number of first stage program executions and 
first stage execution times in microseconds for matrix-matrix multiply and matrix-vector multiply. We
report both BuildIt dynamic control flow exploration executions ~\cite{ajay-thesis} and executions triggered by prophecy variable precondition violations.}
\label{tbl:einsum-reruns}
\vspace{-1.5em}
\end{table}

\subsubsection{Performance Evaluation}
We evaluate the optimized generated code and first-stage overhead of prophecy variables. Without prophecy variables, a DSL designer has four alternatives:
\begin{enumerate}
    \item \textbf{Assume Conservative Analyses Result}: Always use the conservative (lattice-top) analysis result. This preserves correctness but causes unnecessary GPU buffer allocations and copies, whose performance cost we measure.
    \item \textbf{Use Runtime Methods}: Defer decisions to runtime where possible. Here, CUDA Unified Memory~\cite{Harris2017UnifiedMemory} makes memory available to both host and GPU, moving it through page faults. It avoids unnecessary copies but incurs runtime overhead, which we measure.
    \item \textbf{Require Explicit User Annotation}: Require users to annotate tensors needed on the GPU and list those accessed at the start of each \texttt{run\_on\_gpu} call. This avoids runtime overhead but harms usability and risks incorrect annotations. Its runtime performance matches prophecy variables, so we omit this comparison.
    \item \textbf{Explicit Data Flow Analysis on IR}: Build an IR or AST for the whole program and use standard data-flow analysis to propagate tensor uses to allocations and copies. It provides equivalent performance but requires domain experts to implement IR analysis and manipulation, contrary to \buildit's lightweight DSL approach; Since the performance would be the same, we omit this comparison.
\end{enumerate}

We therefore compare prophecy variables with \textit{Copy All} (conservative) and \textit{Unified Memory} (runtime). We measure generated GPU execution time, first-stage execution time, and reexecution counts with and without prophecy variables, asking:
\begin{enumerate}
    \item How much do selective allocations and copies improve matrix-matrix and matrix-vector multiplication over these baselines?
    \item What additional first-stage time and reexecutions are required to obtain correct prophecy-variable analysis results?
\end{enumerate}

Since performance-critical work in \buildit occurs in the second stage, first-stage overhead can be amortized across repeated executions of code generated once.

Tables~\ref{tbl:matrixmatrix} and~\ref{tbl:matrixvector} report minimum GPU execution times in microseconds over ten runs for matrix-matrix and matrix-vector multiplication, respectively, under all three strategies. Experiments use an NVIDIA DGX-1 with an Intel(R) Xeon(R) E5-2698 v4 @ 2.20GHz CPU and a Tesla V100-SXM2-32GB GPU. Prophecy variables consistently perform best, though the performance gap narrows with matrix size: matrix multiplication requires $O(n^3)$ computation but $O(n^2)$ communication, making communication increasingly less significant.

Table~\ref{tbl:einsum-reruns} reports first-stage times and execution counts for both BuildIt's dynamic control flow exploration (ensuring correct second-stage code generation~\cite{buildit, ajay-thesis}) and obtaining correct prophecy-variable values. Prophecy variables can introduce additional paths to explore, increasing dynamic-control-flow executions and first-stage time.

\subsection{Neural Network Operation Fusion}
\label{sec:nn}

\buildit-based DSLs traditionally support operator fusion by deferring an operation so that it can be fused into a later operation. Here, we show how prophecy variables enable future operations to be fused into earlier ones, expanding the opportunities for fusion. Our benchmark uses a machine learning DSL supporting convolutions, fully connected layers, and activation functions such as ReLU. Developers specify networks as sequences of these operations, and BuildIt generates the corresponding second-stage code.

Figure~\ref{lst:mlfusion} shows a convolution followed by a ReLU with an activation threshold. Without fusion, the implementation traverses the tensor once for convolution and again for ReLU. Since ReLU is pointwise, it can instead be fused into the convolution, eliminating the second traversal. Such fusion is an important optimization for neural network performance.

History and prophecy variables work together to enable this fusion in the generated second-stage code. Control flow complicates correctness: a convolution may be followed by ReLU on some paths but not others, and even if all paths have a ReLU, their activation thresholds may differ. Fusion is therefore possible only when every operation immediately following the convolution is a ReLU with the same threshold. The implementation uses:
\begin{itemize}
\item {\bf History Variable:} \bldit{is_last_convolution} tracks whether the last performed tensor operation was a convolution.
\item {\bf Prophecy Variable:} \bldit{is_next_relu} predicts whether all operations immediately following a convolution are ReLUs with the same activation threshold, storing that threshold when they are.
\end{itemize}

All the corresponding logic executes in the first stage. Each operation updates the history variable upon completion; on entry, the next operation checks whether its predecessor was a convolution and, if so, updates the prophecy variable to reflect its operation type and, for ReLU, its threshold. The convolution checks the prophecy variable to generate fused convolution-ReLU code when all immediately following operations are ReLUs with the same threshold, and convolution-only code otherwise. The ReLU also checks the prediction: when fusion is enabled, it generates no second-stage code because it has already been incorporated into the convolution; otherwise it generates the ReLU operation separately.

The prophecy-variable implementation overhead is reasonable: the DSL contains 161 lines of C++ code, 48 of which handle prophecy or history variables. The complete implementation is in supplementary Section~\ref{sec:nnimpl}.

We evaluate the performance benefits and first-stage overhead of prophecy-variable-enabled operator fusion. Specifically, we seek to answer the following research questions:
\begin{enumerate}
    \item How much does fusing convolution and ReLU improve second-stage execution performance compared to executing the two operations separately?
    \item What is the first-stage overhead of enabling this optimization, in terms of additional program reexecutions and execution time?
\end{enumerate}

Tables~\ref{tbl:mlfusion} and~\ref{tbl:convolutionrelu-reruns} present the corresponding results. For the benchmark in Figure~\ref{lst:mlfusion}, Table~\ref{tbl:mlfusion} reports execution times with convolution/ReLU fusion enabled and disabled for 1-D convolutions across input sizes from 10240 to 1024000 and 3- and 9-point filters. Experiments use a 12th Gen Intel(R) Core(TM) i9-12900HK with 32GB of memory. Table~\ref{tbl:convolutionrelu-reruns} reports the first-stage execution counts and times needed to explore all dynamic control flow paths and obtain correct prophecy-variable values. The results show that convolution-ReLU fusion consistently improves second stage performance, with speedups ranging from 1.04x to 1.11x across the evaluated configurations. This improvement comes at the cost of additional first stage execution: prophecy variables require 4 reexecutions, increasing the first stage execution time from 9.9 ms to 14.7 ms (approximately 49\%). However, this overhead is incurred only during code generation and can be amortized over repeated executions of the generated second stage code.

Neural networks typically combine multiple layers with activation functions and other operations such as normalization. Together, history and prophecy variables enable BuildIt to optimize across these layers and operations, fusing them to eliminate unnecessary intermediate temporaries and repeated tensor traversals.

\begin{figure}
\small
\begin{lstlisting}[language=buildit]
void network(void) {
    ml::tensor input(102400, {...});
    ml::tensor weight(9, {...});

    ml::tensor conv_out = ml::convolve(input, weight);
    ml::tensor out = ml::relu(conv_out, 1.35);
    ... // downstream operations 
}
\end{lstlisting}
\caption{\label{lst:mlfusion} BuildIt code for convolution followed by ReLU.}
\end{figure}

\begin{table}
\raggedright   
\begin{minipage}[t]{0.4\textwidth}
\centering
\small
\begin{tabular}{|l|r|r|}
\hline
Image  & Fusion  & Fusion  \\ 
Size & Disabled &  Enabled \\ \hline
10240 $\times$ 3 & 26 & \textbf{25} \\ \hline
102400 $\times$ 3 & 267 & \textbf{243} \\ \hline
1024000 $\times$ 3 & 2,594 & \textbf{2,341} \\ \hline
\hline
10240 $\times$ 9 & 76 & \textbf{73} \\ \hline
102400 $\times$ 9 & 770 & \textbf{742} \\ \hline
1024000 $\times$ 9 & 7,953 & \textbf{7,451} \\ \hline
\end{tabular}
\caption{\label{tbl:mlfusion} Execution times in microseconds for 1-D convolution followed by ReLU with fusion disabled and enabled for different input and filter sizes.}
\end{minipage}
\hfill 
\begin{minipage}[t]{0.56\textwidth}   
\centering
\small
\begin{tabular}{|l|r|r|}
\hline
{Convolution-ReLU} & Fusion & Fusion \\ 
& Disabled & Enabled \\ \hline 
Prophecy Variable Reexecutions & - & 4 \\ \hline
Dynamic Control Flow Executions & 33 &  72 \\ \hline
First Stage Execution Time & 9,872 & 14,701 \\ \hline
\end{tabular}
\caption{Number of first stage program executions and 
first stage execution times in microseconds. We
report both reexecutions triggered by prophecy variable precondition violations and dynamic control flow exploration executions~\cite{ajay-thesis}.}
\label{tbl:convolutionrelu-reruns}
\end{minipage}
\vspace{-1em}
\end{table}

mimitmi%

\subsection{Binding Time Analysis}
\label{sec:binding}

One of the biggest cognitive hurdles for users to implement \buildit DSLs is having to reason about whether variables are static or dynamic. In this section, we will describe a prophecy variable based binding time analysis that completely eliminates this.

In traditional \buildit programs, developers explicitly declare variables as either \bldit{static_var<T>} or \bldit{dyn_var<T>}, determining whether their operations execute in the first or second stage. This presents two challenges:

\begin{itemize}
    \item {\bf Manual Binding-Time Decisions:} Even when the distinction between first-stage and second-stage inputs is fixed, developers must reason about how these values propagate through the DSL implementation to determine which variables should be static and which must be dynamic. This adds significant cognitive overhead to DSL development.

    \item {\bf Reusing Abstractions Across Contexts:} The same abstraction may be used in different contexts, where a variable needs to be static in one context and dynamic in another. Since these types are fixed in the implementation, supporting both contexts may require rewriting or duplicating the abstraction.
\end{itemize}

To address both challenges, we introduce \bldit{union_var<T>}, which automatically resolves to either \bldit{static_var<T>} or \bldit{dyn_var<T>} based on how the variable is used throughout the program. This removes the need to manually identify binding times and allows the same abstraction to adapt to different combinations of static and dynamic variables. The analysis also supports gradual typing, allowing \bldit{union_var<T>}, \bldit{static_var<T>}, and \bldit{dyn_var<T>} to be freely intermixed.

The challenge is that determining whether a variable can be evaluated in the first stage requires information about its future use. A variable may initially receive only static values but later be assigned a dynamic value, requiring it to be treated as dynamic from the beginning. We use prophecy variables to predict whether each \bldit{union_var<T>} must be dynamic, with reexecution correcting predictions when necessary.

Specifically, each \bldit{union_var<T>} contains a \bldit{static_var<T>} and a \bldit{dyn_var<T>} component (see Section~\ref{sec:union}). A prophecy variable predicts which component will be used, with dynamic above static in the lattice. The initial prediction is static unless the variable is initialized with a dynamic value. Reads access the predicted component, while assignments, implemented by the \bldit{union_var<T>} assignment operator, assert that the prediction is at least static or dynamic according to the assigned value. A violated assertion causes reexecution with an updated prediction. Thus, static variables are represented by first-stage values, while dynamic variables generate a single second-stage variable of type \bldit{T}. Section~\ref{sec:union} of the supplementary material lists the complete implementation. 

An important interaction arises with first-stage control flow exploration. If a loop iterator of type \bldit{union_var<T>} is predicted static, the first stage unrolls the loop, assigning a new static value at each iteration. To limit excessive unrolling and second-stage code generation, the prophecy-variable lattice also tracks the number of first-stage assignments. Its static elements are ordered by assignment count, with fewer assignments below more; $\tuple{\mbox{static}, 0}$ is the bottom element and $\tuple{\mbox{dynamic}}$ is the top. Once the assignment count exceeds a threshold, the variable is predicted dynamic. This resembles lattice widening~\cite{CousotCousot1977,RivalYi2020}, but controls generated code size rather than dataflow analysis convergence.

We evaluate the first-stage overhead of automatically determining variable binding times using prophecy variables. Specifically, we seek to answer: How many additional first-stage program executions are required, and how much execution time is incurred, when variables are automatically resolved to static or dynamic compared to explicitly declaring their binding times? Figure~\ref{fig:binding} presents these results for both static and dynamic resolutions of tensor shapes and sizes, compared against a baseline with explicitly static shapes and sizes and dynamic tensor values.
The first uses explicit \bldit{static_var<T>} tensor shapes and sizes, with dynamic tensor values, as in Section~\ref{sec:einsum}. The other two use the same tensor implementation but declare shape and size variables as \bldit{union_var<T>}. In the second version, the application context resolves these variables to static, specializing the generated code to the shapes and sizes; in the third, it resolves them to dynamic, permitting shapes and sizes to vary in the second stage. Each tensor has multiple shape and size variables and therefore multiple prophecy variables. All three versions also use prophecy and history variables for CPU/GPU tensor movement as in Section~\ref{sec:einsum}.

For each version, Figure~\ref{fig:binding} reports prophecy-variable reexecutions caused by violated preconditions, executions for dynamic control flow exploration, and first-stage execution time. The first version's results match the last column of Table~\ref{tbl:einsum-reruns}. Reexecutions increase from the first to the second version because loop iterators declared as \bldit{union_var<T>} must change from predicted static to dynamic when they depend on the dynamic thread ID. They increase further in the third version because tensor shapes and sizes also change from initially static predictions to dynamic, triggering additional precondition violations. Each prophecy-variable reexecution entails further dynamic control flow exploration, increasing both execution counts and first-stage time which introduces significant overhead. 
However, this overhead is confined to the first stage and can be amortized over repeated second stage executions. In exchange, the analysis eliminates the need for developers to manually determine variable binding times and enables the same DSL abstractions to be reused across different static and dynamic contexts. Furthermore, the first stage can also report the automatically inferred static or dynamic types to developers, who may then replace \bldit{union_var<T>} with explicit \bldit{static_var<T>} or \bldit{dyn_var<T>} declarations if they wish to make these distinctions visible in the code and eliminate the analysis overhead on consecutive runs.

\begin{figure}
\centering
\footnotesize
\begin{tabular}{|l|R{2.2cm}|R{3.4cm}|R{3.1cm}|}
\hline
{Matrix-Matrix Multiply} & \blditsmall{static_var<T>} shapes and sizes & \blditsmall{union_var<T>} shapes and sizes resolved to \blditsmall{static_var<T>} & \blditsmall{union_var<T>} shapes and sizes resolved to \blditsmall{dyn_var<T>} \\ \hline 
Prophecy Variable Reexecutions & 7 & 17 & 98 \\ \hline
Control Flow Executions & 113 & 250 & 482 \\ \hline
First Stage Execution Time & 38,335 & 336,999 & 752,544 \\ \hline \noalign{\vskip 8pt} \hline
{Matrix-Vector Multiply} & \blditsmall{static_var<T>} shapes and sizes & \blditsmall{union_var<T>} shapes and sizes resolved to \blditsmall{static_var<T>} & \blditsmall{union_var<T>} shapes and sizes resolved to \blditsmall{dyn_var<T>} \\ \hline 
Prophecy Variable Reexecutions & 7 & 10 & 46 \\ \hline
Control Flow Executions & 89 & 128 &  247 \\ \hline
First Stage Execution Time &  27,167 & 159,268 & 339,795 \\ \hline 
\end{tabular}
\caption{\label{fig:binding} Number of first stage program executions and 
first stage execution times in microseconds for matrix-matrix multiply and matrix-vector multiply. We
report both BuildIt dynamic control flow exploration executions ~\cite{ajay-thesis} and executions triggered by prophecy variable precondition violations. We compare the implementation without the binding time analysis (developer specified \blditsmall{static_var<T>} shapes and sizes), with binding time analysis where the shapes and sizes are declared \blditsmall{union_var<T>} and resolve to \blditsmall{static_var<T>} (but loop iterators resolve to \blditsmall{dyn_var<T>}), and with binding time analysis where \blditsmall{union_var<T>} shapes, sizes, and loop iterators all resolve to \blditsmall{dyn_var<T>}. }
\label{tbl:einsum-reruns-binding}
\vspace{-1em}
\end{figure}

	\section{Related Work}
\label{sec:related}

We review prophecy and history variables, DSL implementations, multistage programming, and binding-time analysis. To our knowledge, this is the first use of prophecy variables for program analysis in a programming language implementation, and the first use of program reexecution to compute correct prophecy-variable values.

\subsection{Prophecy and History Variables}

History variables were first proposed for program verification~\cite{OwickiG76}. Prophecy variables were first introduced to enable refinement-mapping proofs with forward simulations~\cite{AbadiL91}. Combined with history variables, they enabled a completeness result for proving trace inclusion~\cite{AbadiL91}. They have also been used for program verification~\cite{JungLPRTDJ20,Vafeiadis08,ZhangFFSL12}. Our goal instead is to enable optimizations in lightweight DSL systems using only forward execution, without an explicit IR.

Two papers, neither published in a peer-reviewed venue, use them for program analysis, but compute their values via backward dataflow analysis over explicit program representations. One specifies live variables and very busy expressions~\cite{rinard2020dataflow}; the other adds subset-inclusion constraints to operational-semantics step rules and uses a backwards dataflow analysis to compute prophecy variable values to verify partial dead code elimination and lazy code motion~\cite{Nexis}. In contrast, we obtain correct predictions through repeated forward execution in BuildIt, allowing its first stage to work with information about the future execution
without an explicit IR and without backwards analyses that operate on that (nonexistent) IR. 

Cobalt~\cite{LernerMC03} specifies optimizations such as constant propagation and partial dead assignment elimination using transformation patterns guarded by conditions on sequences of actions along program paths, with separate forward and backward guards. Rhodium~\cite{LernerMRC05} avoids temporal-logic formulas and implicit dataflow facts by relating concrete states to dataflow facts through abstraction-function predicates and using state extensions (instrumented semantics) to capture past-execution information. Both support forward and backward analyses; later work inferring propagation rules automatically supports only forward rules~\cite{ScherpelzLC07}.

\subsection{Domain Specific Language Implementations}

Many DSLs emphasize productivity rather than performance, including graphics languages TikZ~\cite{tikz}, Vega/Vega-Lite~\cite{vega,vegalite}, gnuplot~\cite{gnuplot}, matplotlib~\cite{matplotlib}, and Graphviz~\cite{graphviz}; build tools make~\cite{make}, cmake~\cite{cmake}, and Ninja~\cite{ninja}; and lexer/parser generators flex/lex and bison/yacc~\cite{flex,yacc}, and ANTLR~\cite{antlr}. Such systems typically require little optimization-oriented program analysis.

Performance-oriented DSLs, including GraphIt~\cite{graphit,g2,ugf}, TACO~\cite{taco}, Halide~\cite{halide}, Diderot~\cite{diderot}, Taichi~\cite{taichi}, and SQL systems~\cite{mysql,sqlite}, commonly use parsers, IRs, and forward and backward analyses to optimize generated code. BuildIt avoids these components; prophecy variables and reexecution enable optimizations requiring future-execution information using only forward execution, without an explicit IR.

\subsection{Partial Evaluation/Multistage Programming}

Partial evaluation and multistage programming have a long history~\cite{Futamura1971,JonesGomardSestoft1993,ConselDanvy1993,DaviesPfenning1996,TahaSheard2000,CaretteKiselyovShan2009,RompfOdersky2010,Danvy1996,Taha2004}. General-purpose staging systems include Template Haskell~\cite{sheard2002template}, Lightweight Modular Staging in Scala~\cite{lms}, Scheme~\cite{scheme}, MetaML~\cite{TahaSheard2000}, Meta OCaml, Mint~\cite{mint}, Terra (in Lua)~\cite{terra,terra2}, C macros, and C++ templates~\cite{vandevoorde2002c++}. Staging also supports embedded DSLs such as PyTorch~\cite{pytorch}, TensorFlow~\cite{tensorflow}, TVM~\cite{Chen2018TVM}, AnyDSL~\cite{anydsl}, Delite~\cite{delite}, LMS~\cite{lms}, BuilDSL~\cite{buildsl}, JAX~\cite{jax2018github}, SHiM~\cite{shim}, and BREeze~\cite{tamara:meng-thesis:2023}.

These systems often use sophisticated analyses implemented conventionally or through history variables or equivalents; to our knowledge, none uses prophecy variables or equivalents. In contrast, BuildIt needs no explicit IR, and prophecy variables with repeated first-stage execution enable future-dependent analyses and optimizations using a standard language with only forward execution.

\subsection{Binding Time Analysis}

Binding-time analysis determines which stage evaluates each expression~\cite{nielsonandnielson,temporallogic,JonesGomardSestoft1993,taha1997multistage,TahaSheard2000}, using type-based~\cite{nielsonandnielson,taha1997multistage,TahaSheard2000} or analysis-based~\cite{JonesGomardSestoft1993} techniques. BuildIt's \bldit{static_var<T>} and \bldit{dyn_var<T>} traditionally require programmers to identify binding times explicitly. Our \bldit{union_var<T>} instead uses prophecy variables to resolve to either type based on whole-program usage, including future execution. It integrates with both existing types in a gradual typing system that allows all three to be freely intermixed.

	\section{Conclusion}
\label{sec:conclusion}

The staged BuildIt lightweight DSL implementation system delivers a combination of substantially reduced implementation effort and performance that is comparable to and often exceeds the performance of heavyweight standard DSL implementations. BuildIt is implemented as C++ data structures that encapsulate sophisticated program information extraction, transformation, and multioperation optimization techniques. This approach eliminates otherwise required program language implementation components such as parsers and intermediate representations. Instead of using dataflow analysis to extract information, BuildIt uses history variables to record and propagate information about the past execution of the program to BuildIt first stage program transformation and optimization sites. 

We introduce the combination of prophecy variables and repeated program execution, with the repeated program execution producing correct prophecy variable predictions. Our results illustrate how this combination extends 
the scope of the BuildIt implementation to include transformations and optimizations that require information about the future program execution while 
preserving the substantial engineering effort advantages of the basic BuildIt approach. By extending the scope of BuildIt to include languages whose efficient implementation 
requires these kinds of transformations and optimizations, 
the presented techniques promise to further eliminate barriers to rapid and efficient DSL implementation, enabling more developers to benefit from these effective application implementation platforms.

	\bibliography{paper.bib}
    \clearpage
    \section*{Supplemental Material for \titletext}

\subsection{Einsum DSL Complete Implementation}
\label{sec:appeinsum}
This supplementary material provides the complete source code for the einsum DSL implemented on top of \buildit. This implementation uses a combination of prophecy variables and history variables to identify tensors to be moved between CPU and GPU. The implementation only has the \buildit multistage library as a dependency and is compiled as regular C++ code. The source code below also includes the DSL program for matrix multiplication and a main function to generate code for the DSL. 
\\
\begin{lstlisting}[language=buildit, escapechar=|]
#include "builder/dyn_var.h"
#include "builder/static_var.h"
#include "builder/generics.h"
#include "builder/nd_var.h"
#include "blocks/extract_cuda.h"
#include "builder/builder_context.h"
#include "blocks/c_code_generator.h"
#include "builder/signature_extract.h"


namespace el {
template <typename T>
using prophecy_var = builder::nd_var<T>;
using builder::with_name;

namespace runtime {
builder::dyn_var<(void*, int)> malloc = with_name("runtime::typed_malloc");
builder::dyn_var<void(void*)> free = with_name("runtime::typed_free");
builder::dyn_var<void(void)> start_time = with_name("runtime::start_time");
builder::dyn_var<void(void)> end_time = with_name("runtime::end_time");
builder::dyn_var<void(void)> grid_sync = with_name("runtime::grid_sync");
builder::dyn_var<void(void*, int)> cuda_malloc = with_name("runtime::cuda_malloc");
builder::dyn_var<void(void*, int)> unified_malloc = with_name("runtime::unified_malloc");
builder::dyn_var<void(void*, void*, int)> cudaMemcpyToDevice = with_name("runtime::cudaMemcpyToDevice");
builder::dyn_var<void(void*, void*, int)> cudaMemcpyToHost = with_name("runtime::cudaMemcpyToHost");
}

static constexpr const int max_bid = 40;
static constexpr const int max_tid = 512;

enum device_type {
        device_cpu,
        device_gpu
};
builder::dyn_var<int> *current_bid = nullptr;
builder::dyn_var<int> *current_tid = nullptr;
device_type current_device = device_cpu;

bool option_copy_all_tensors = false;
bool option_use_unified_memory = false;

struct index;
struct tensor_base;
template <typename T> struct tensor;
template <typename T> struct tensor_access;
template <typename T> struct tensor_term;

struct index {
	builder::dyn_var<int> *m_it;
	// Index ranges are decided when they are used
	int m_range = -1;
};
|\clearpage|
struct tensor_base {
	static std::vector<tensor_base*> registered_tensors;
    
	prophecy_var<builder::true_top> needs_gpu = builder::true_top::F;

	builder::static_var<bool> gpu_written = false;
	prophecy_var<builder::true_top> *gpu_read = nullptr;

	tensor_base() {
		registered_tensors.push_back(this);
	}
    
	virtual ~tensor_base();
	
	virtual void move_to_gpu() = 0;
	virtual void move_to_host() = 0;
};

// Type declarations
template <typename T>
struct tensor: public tensor_base {
	std::vector<int> m_sizes;
	int m_elemsize;
	
	builder::dyn_var<T*> m_buffer;
	builder::dyn_var<T*> m_gpu_buffer;

	tensor(const std::vector<int> &sizes): m_sizes(sizes), m_elemsize(sizeof(T)) {
		if (option_use_unified_memory) {	
			runtime::unified_malloc(&m_buffer, get_total_size() * m_elemsize);	
		} else {
			runtime::malloc(&m_buffer, get_total_size() * m_elemsize);	
			if (option_copy_all_tensors or needs_gpu.get() == builder::true_top::T) {
				runtime::cuda_malloc(&m_gpu_buffer, get_total_size() * m_elemsize);
			}
		}	
	}

	tensor(const std::vector<int> &sizes, const builder::dyn_var<T*>& buffer): m_sizes(sizes), 
            m_elemsize(sizeof(T)) {
		// if user supplies buffer, it is users responsibility to use unified memory if necessary
		if (!option_use_unified_memory) {
			if (option_copy_all_tensors or needs_gpu.get() == builder::true_top::T) {
				runtime::cuda_malloc(&m_gpu_buffer, get_total_size() * m_elemsize);
			}
		}
	}
	~tensor() {}

	int get_total_size();

	// Delete the copy constructors and assignments
	tensor(const tensor&) = delete;
	tensor(tensor&&) = delete;
	tensor& operator=(const tensor&) = delete;
	tensor& operator=(tensor&&) = delete;
		
	// operations	
	tensor_access<T> operator[] (index& i);

	void move_to_gpu() override;
	void move_to_host() override;
};
|\clearpage|
template <typename T>
struct tensor_access {
	tensor<T>& m_tensor;
	std::vector<index*> m_indices;
	
	tensor_access(tensor<T>& t, index& i): m_tensor(t), m_indices({&i}) {}

	void operator=(const tensor_term<T>&);
	void operator+=(const tensor_term<T>&);
	void operator*=(const tensor_term<T>&);

	void operator=(const tensor_access<T>& o);
	
	tensor_access<T> operator[] (index& i) const;

	std::vector<index*> gather_indices(void) const;
	void gather_ranges(void) const;

	std::pair<std::vector<index*>, std::vector<index*>> initialize_indices(const tensor_term<T>&rhs);

	enum reduction_op {red_assign, red_sum, red_product};
	void induce_lhs_loops(const std::vector<index*> &lhs_indices, int ind, 
		enum reduction_op op, const std::vector<index*> &rhs_indices, const tensor_term<T>& rhs) const;
	void induce_rhs_loops(const std::vector<index*> &rhs_indices, int ind, reduction_op op, 
		const tensor_term<T>& rhs, builder::dyn_var<T>* agg) const;

	builder::dyn_var<int> get_flat_index(int index) const;

};

template <typename T>
struct tensor_term {
	enum mode {
		mode_tensor, 
		mode_value,
		mode_sum, 
		mode_product,
		mode_index, 
	};
	mode m_mode;
	const tensor_access<T>* m_tensor;
	// allow the m_value to be mutable because the type is set later
	builder::dyn_var<T> m_value;
	std::unique_ptr<const tensor_term<T>> m_e1;
	std::unique_ptr<const tensor_term<T>> m_e2;
	index* m_index;

	// The 4 constructors, sum and product share a constructor
	// The builder with_name trick is used to suppress the construction of m_value when not required
	// We could also used defer_init
	tensor_term(const tensor_access<T>& ta): m_mode(mode_tensor), m_tensor(&ta),
		m_value(builder::with_name("__discard")) {}	
	tensor_term(tensor_term&& e1, tensor_term&& e2): 
		m_e1(std::unique_ptr<tensor_term<T>>(new tensor_term<T>(std::move(e1)))), 
		m_e2(std::unique_ptr<tensor_term<T>>(new tensor_term<T>(std::move(e2)))), 
		m_value(builder::with_name("__discard")) {
		}
	tensor_term(index &i): m_mode(mode_index), m_index(&i),
		m_value(builder::with_name("__discard")) {}	

	// A catch all constructor for anything that is convertible to dyn_var
	template <typename T2>
	tensor_term (const T2& t): m_mode(mode_value), m_value(t) {}
	
	std::vector<index*> gather_indices(void) const;
	void gather_ranges(void) const;

	builder::dyn_var<T> get_value() const;	
};
|\clearpage|
// Type method definitions
template <typename T>
tensor_access<T> tensor<T>::operator[] (index &i) {
	return tensor_access(*this, i);
}
template <typename T>
int tensor<T>::get_total_size(void) {
	int total = 1;
	for (auto s: m_sizes) {
		total *= s;
	}
	return total;
}

template <typename T>
tensor_access<T> tensor_access<T>::operator[] (index& i) const {
	tensor_access ret (*this);
	ret.m_indices.push_back(&i);
	return ret;
}
template <typename T>
std::pair<std::vector<index*>, std::vector<index*>> 
tensor_access<T>::initialize_indices(const tensor_term<T>&rhs) {
	// Initialize indices for acquiring range
	std::vector<index*> lhs_indices = gather_indices();
	std::vector<index*> rhs_indices = rhs.gather_indices();
	for (auto a: lhs_indices) {
		a->m_range = -1;
	}
	for (auto a: rhs_indices) {
		a->m_range = -1;
	}

	gather_ranges();	
	rhs.gather_ranges();
	
	// At this point make sure all indices used have ranges 
	for (auto a: lhs_indices) {
		assert(a->m_range != -1);
	}
	for (auto a: rhs_indices) {
		assert(a->m_range != -1);
	}

	std::vector<index*> reduce_indices;
	for (auto a: rhs_indices) {
		if (std::find(lhs_indices.begin(), lhs_indices.end(), a) == lhs_indices.end()) {
			reduce_indices.push_back(a);
		}
	}
	return std::make_pair(lhs_indices, reduce_indices);
}

template <typename T>
builder::dyn_var<int> tensor_access<T>::get_flat_index(int ind) const {
	if ((ind + 1) == m_indices.size()) return *(m_indices[ind]->m_it);
	int size_after = 1;
	for (unsigned i = ind + 1; i < m_tensor.m_sizes.size(); i++) {
		size_after *= m_tensor.m_sizes[i];
	}
	return (builder::cast)(size_after * *(m_indices[ind]->m_it) + get_flat_index(ind + 1));
}
|\clearpage|
template <typename T>
void tensor_access<T>::induce_rhs_loops(const std::vector<index*> &rhs_indices, int ind, reduction_op op, 
		const tensor_term<T>& rhs, builder::dyn_var<T>* agg) const {
	if (ind == rhs_indices.size()) {
		if (op == red_product) {
			*agg = *agg * rhs.get_value();
		} else if (op == red_sum) {
			*agg = *agg + rhs.get_value();
		}
		return;
	}
	for (builder::dyn_var<int> i = 0; i < rhs_indices[ind]->m_range; ++i) {
		rhs_indices[ind]->m_it = i.addr();
		induce_rhs_loops(rhs_indices, ind + 1, op, rhs, agg);
	}
	rhs_indices[ind]->m_it = nullptr;
}

template <typename T>
void tensor_access<T>::induce_lhs_loops(const std::vector<index*> &lhs_indices, int ind, reduction_op op, 
		const std::vector<index*> &rhs_indices, const tensor_term<T>& rhs) const {
	if (ind == lhs_indices.size()) {
		builder::dyn_var<T*>* update_var;
		if (current_device == device_gpu && !option_use_unified_memory) {
			if (!option_copy_all_tensors)
				m_tensor.needs_gpu.assert_requires_atleast(builder::true_top::T);
			update_var = m_tensor.m_gpu_buffer.addr();
			if (!option_copy_all_tensors && !option_use_unified_memory)
				m_tensor.gpu_written = true;
		} else {
			update_var = m_tensor.m_buffer.addr();
		}
		if (op == red_assign) {
			(*update_var)[get_flat_index(0)] = rhs.get_value();
		} else if (op == red_product) {
			builder::dyn_var<T> prod = 1;
			induce_rhs_loops(rhs_indices, 0, op, rhs, prod.addr());	
			(*update_var)[get_flat_index(0)] = prod;	
		} else if (op == red_sum) {
			builder::dyn_var<T> sum = 0;
			induce_rhs_loops(rhs_indices, 0, op, rhs, sum.addr());	
			(*update_var)[get_flat_index(0)] = sum;
		}
		return;
	}
	// GPU Specializations
	if (current_device == device_gpu) {
		// If the lhs has only one loop, use the TID directly
		if (lhs_indices.size() == 1) {
			builder::dyn_var<int> thread = (*current_bid) * max_tid + (*current_tid);
			for (builder::dyn_var<int> i = thread; i < lhs_indices[ind]->m_range; i += max_tid * max_bid) {
				lhs_indices[ind]->m_it = i.addr();	
				induce_lhs_loops(lhs_indices, ind + 1, op, rhs_indices, rhs);
			}
			lhs_indices[ind]->m_it = nullptr;
			runtime::grid_sync();
			return;
		} else {
			// Special case for index 1 and 2	
			if (ind == 0) {
				for (builder::dyn_var<int> i = *current_bid; i < lhs_indices[0]->m_range; i += max_bid) {
					lhs_indices[0]->m_it = i.addr();	
					for (builder::dyn_var<int> j = *current_tid; j < lhs_indices[1]->m_range; j+= max_tid) {
						lhs_indices[1]->m_it = j.addr();	
						induce_lhs_loops(lhs_indices, ind + 2, op, rhs_indices, rhs);
					}
				}
				lhs_indices[0]->m_it = nullptr;
				lhs_indices[1]->m_it = nullptr;
				runtime::grid_sync();
				return;	
			}
		}
	}

	// to allow parallelization we have to use a copy
	for (builder::dyn_var<int> i = 0; i < lhs_indices[ind]->m_range; ++i) {
		lhs_indices[ind]->m_it = i.addr();
		induce_lhs_loops(lhs_indices, ind + 1, op, rhs_indices, rhs);
	}
	lhs_indices[ind]->m_it = nullptr;
}

template <typename T>
void tensor_access<T>::operator=(const tensor_term<T>& rhs) {
	auto indices = initialize_indices(rhs);	
	std::vector<index*> lhs_indices = indices.first;
	std::vector<index*> reduce_indices = indices.second;

	// Specifically for assignment, reduce_indices should be empty		
	assert(reduce_indices.size() == 0);

	induce_lhs_loops(lhs_indices, 0, red_assign, reduce_indices, rhs);	
}

template <typename T>
void tensor_access<T>::operator=(const tensor_access<T>& o) {
	this->operator=((tensor_term<T>)o);
}

template <typename T>
void tensor_access<T>::operator+=(const tensor_term<T>& rhs) {
	auto indices = initialize_indices(rhs);	
	std::vector<index*> lhs_indices = indices.first;
	std::vector<index*> reduce_indices = indices.second;
	induce_lhs_loops(lhs_indices, 0, red_sum, reduce_indices, rhs);	
}

template <typename T>
void tensor_access<T>::operator*=(const tensor_term<T>& rhs) {
	auto indices = initialize_indices(rhs);	
	std::vector<index*> lhs_indices = indices.first;
	std::vector<index*> reduce_indices = indices.second;
	induce_lhs_loops(lhs_indices, 0, red_product, reduce_indices, rhs);	
}

template <typename T>
std::vector<index*> tensor_access<T>::gather_indices(void) const {
	// Make sure all indices are provided
	assert(m_indices.size() == m_tensor.m_sizes.size());
	// deduplicate indices and return
	std::vector<index*> ret;
	for (auto a: m_indices) {
		if (std::find(ret.begin(), ret.end(), a) == ret.end()) {
			ret.push_back(a);
		}
	}
	return ret;
}

template <typename T>
void tensor_access<T>::gather_ranges(void) const {
	for (unsigned int i = 0; i < m_indices.size(); i++) {
		auto ind = m_indices[i];
		if (ind->m_range != -1) {
			assert(ind->m_range == m_tensor.m_sizes[i]);
		}
		ind->m_range = m_tensor.m_sizes[i];
	} 
}
|\clearpage|
template <typename T>
std::vector<index*> tensor_term<T>::gather_indices(void) const {
	if (m_mode == mode_value) {
		return {};
	} else if (m_mode == mode_tensor) {
		return m_tensor->gather_indices();
	} else if (m_mode == mode_product or m_mode == mode_sum) {
		std::vector<index*> ret = m_e1->gather_indices();
		std::vector<index*> other = m_e2->gather_indices();
		for (auto a: other) {
			if (std::find(ret.begin(), ret.end(), a) == ret.end()) {
				ret.push_back(a);
			}
		}	
		return ret;
	} else if (m_mode == mode_index) {
		return {m_index};
	}
	return {};
}

template <typename T>
void tensor_term<T>::gather_ranges(void) const {
	if (m_mode == mode_tensor) {
		m_tensor->gather_ranges();
	} else if (m_mode == mode_product or m_mode == mode_sum) {
		m_e1->gather_ranges();
		m_e2->gather_ranges();
	}
	// index doesn't gather ranges
}

template <typename T>
builder::dyn_var<T> tensor_term<T>::get_value(void) const {
	if (m_mode == mode_value) {
		return m_value;	
	} else if (m_mode == mode_tensor) {
		if (current_device == device_gpu && !option_use_unified_memory) {
			if (!option_copy_all_tensors)
				m_tensor->m_tensor.needs_gpu.assert_requires_atleast(builder::true_top::T);
			if (!option_copy_all_tensors && !option_use_unified_memory)
				m_tensor->m_tensor.gpu_read->assert_requires_atleast(builder::true_top::T);
			return m_tensor->m_tensor.m_gpu_buffer[m_tensor->get_flat_index(0)];
		} else {
			return m_tensor->m_tensor.m_buffer[m_tensor->get_flat_index(0)];
		}
	} else if (m_mode == mode_product) {
		return m_e1->get_value() * m_e2->get_value();
	} else if (m_mode == mode_sum) {
		return m_e1->get_value() + m_e2->get_value();
	} else if (m_mode == mode_index) {
		return *(m_index->m_it);
	}
	return 0;
}
template <typename T>
void tensor<T>::move_to_gpu() {
	runtime::cudaMemcpyToDevice(m_gpu_buffer, m_buffer, get_total_size() * m_elemsize);
}
template <typename T>
void tensor<T>::move_to_host() {
	runtime::cudaMemcpyToHost(m_buffer, m_gpu_buffer, get_total_size() * m_elemsize);
}
|\clearpage|
// utility SFINAE to check if a type is
// one of the types that can appear in expressions

template <typename T>
struct is_tensor_expr: std::false_type {};
template <typename T>
struct is_tensor_expr<tensor_term<T>>: std::true_type {
        using wrapped_type = T;
};
template <typename T>
struct is_tensor_expr<tensor_access<T>>: std::true_type {
        using wrapped_type = T;
};

template <typename T1, typename T2, typename V=void>
struct is_valid_arg_types {};

template <typename T1, typename T2>
struct is_valid_arg_types<T1, T2, typename std::enable_if<is_tensor_expr<T1>::value>::type> {
        using type = tensor_term<typename is_tensor_expr<T1>::wrapped_type>;
};
template <typename T1, typename T2>
struct is_valid_arg_types<T1, T2, typename std::enable_if<!is_tensor_expr<T1>::value && 
            is_tensor_expr<T2>::value>::type> {
        using type = tensor_term<typename is_tensor_expr<T2>::wrapped_type>;
};


template <typename T1, typename T2>
auto operator + (T1&& e1, T2&& e2) -> typename is_valid_arg_types<T1, T2>::type {
        using tt = typename is_valid_arg_types<T1, T2>::type;
        tt ret(e1, e2);
        ret.m_mode = tt::mode_sum;
        return ret;
}
template <typename T1, typename T2>
auto operator * (T1&& e1, T2&& e2) -> typename is_valid_arg_types<T1, T2>::type {
        using tt = typename is_valid_arg_types<T1, T2>::type;
        tt ret(std::move(e1), std::move(e2));
        ret.m_mode = tt::mode_product;
        return ret;
}

template <typename RetType, typename ArgTuple>
struct wrap_helper;

template <typename RetType, typename...Args>
struct wrap_helper<RetType, std::tuple<Args...>> {
        static std::function<RetType(Args...)> wrap (std::function<RetType(Args...)> f) {
                return [=](Args...args) -> RetType {
                        current_device = device_cpu;
                        return f(args...);
                };
        }
};

template <typename...Args>
struct wrap_helper<void, std::tuple<Args...>> {
        static std::function<void(Args...)> wrap (std::function<void(Args...)> f) {
                return [=](Args...args) -> void {
                        current_device = device_cpu;
                        f(args...);
                };
        }
};

|\clearpage|

void run_on_gpu(std::function<void(void)> func);
void run_on_gpu(std::function<void(void)> func) {
	current_device = device_gpu;

	// Before launching the kernel, non-deterministically guess which 
	// tensors are required on the gpu

	for (builder::static_var<unsigned> i = 0; i < tensor_base::registered_tensors.size(); i++) {
		auto t = tensor_base::registered_tensors[i];
		if (!option_use_unified_memory)	{
			if (option_copy_all_tensors) {
				t->move_to_gpu();
			} else {
				
				if (t->gpu_read) delete t->gpu_read;
                t->gpu_written = false;
				t->gpu_read = new prophecy_var<builder::true_top>(builder::true_top::F);
				

				if (t->gpu_read->get() == builder::true_top::T) {
					t->move_to_gpu();
				}
			}
		}
	}

	builder::annotate(CUDA_KERNEL_COOP);
	for (builder::dyn_var<int> bid = 0; bid < max_bid; ++bid) {
		for (builder::dyn_var<int> tid = 0; tid < max_tid; ++tid) {
			current_bid = bid.addr();
			current_tid = tid.addr();
			func();
		}
	}
	current_bid = nullptr;
	current_tid = nullptr;
	current_device = device_cpu;

	// After finishing copy back tensors that have been written and delete the nd vars

	for (builder::static_var<unsigned> i = 0; i < tensor_base::registered_tensors.size(); i++) {
		auto t = tensor_base::registered_tensors[i];

		if (!option_use_unified_memory) {
			if (option_copy_all_tensors) {
				t->move_to_host();
			} else {
				if (t->gpu_written) {
					t->move_to_host();
				}

				delete t->gpu_read;
				t->gpu_written = false;
				t->gpu_read = nullptr;
			}
		}
	}
}
|\clearpage|
template <typename T, typename...Args>
void generate_code(std::ostream& oss, const T& func, std::string fname, Args&&...args) {
        builder::builder_context ctx;
        ctx.run_rce = true;
        using ArgTypes = typename builder::f_type_helper<T>::arg_types;
        using ReturnType = typename builder::f_type_helper<T>::ret_type;
        auto wrapped = wrap_helper<ReturnType, ArgTypes>::wrap(func);
        auto ast = ctx.extract_function_ast(wrapped, fname, std::forward<Args>(args)...);

        auto new_decls = block::extract_cuda_from(block::to<block::func_decl>(ast)->body);

        oss << "#include \"el_runtime.h\"" << std::endl;
        for (auto a: new_decls)
                block::c_code_generator::generate_code(a, oss, 0);
        block::c_code_generator::generate_code(ast, oss, 0);
}

std::vector<tensor_base*> tensor_base::registered_tensors;

tensor_base::~tensor_base() {
        registered_tensors.erase(std::remove(registered_tensors.begin(), registered_tensors.end(), this),
                registered_tensors.end());
        if (gpu_read) delete gpu_read;
}

}

void benchmark (builder::dyn_var<float*> _x, builder::dyn_var<float*> _y, 
    builder::dyn_var<float*> _z, int M, int N, int O) {
        el::index i, j, k;

        el::tensor<float> x({M, N});
        el::tensor<float> x_b({M, N}, _x);
        el::tensor<float> y({N, O});
        el::tensor<float> y_b({N, O}, _y);
        el::tensor<float> z({M, O});


        // temporary to read data into
        el::tensor<float> z_b({M, O}, _z);


        for (builder::dyn_var<int> iter = 0; iter < 10; iter++) {

                x[i][j] = x_b[i][j];
                y[i][j] = y_b[i][j];

                el::runtime::start_time();

                el::run_on_gpu([&]() {
                        z[i][j] += x[i][k] * y[k][j];
                });

                z_b[i][j] = z[i][j];

                el::runtime::end_time();
        }

}


int main(int argc, char* argv[]) {
        el::generate_code(std::cout, benchmark, "benchmark", 1024, 1024, 1024);
        el::option_copy_all_tensors = true;
        el::generate_code(std::cout, benchmark, "benchmark_copy_all", 1024, 1024, 1024);
        el::option_use_unified_memory = true;
        el::generate_code(std::cout, benchmark, "benchmark_unified", 1024, 1024, 1024);
        return 0;
}
\end{lstlisting}

\clearpage

\subsection{Neural Network Fusion DSL Complete Implementation}
\label{sec:nnimpl}
This supplementary material provides the complete source code for the mini neural network DSL implemented on top of \buildit. This implementation uses a combination of prophecy variables and history variables to automatically fuse nonlinearities into complex operations like convolutions. The implementation only has the \buildit multistage library as a dependency and is compiled as regular C++ code. The source code below also includes the DSL program for simple test where the operations cannot be fused due to ReLU with different thresholds on different paths and an example where it can be used. 
\\
\begin{lstlisting}[language=buildit,escapechar=|]
#include "builder/dyn_var.h"
#include "builder/static_var.h"
#include "builder/builder_context.h"
#include "blocks/c_code_generator.h"
#include "builder/nd_var.h"

namespace ml {
template <typename T>
using prophecy_var = builder::nd_var<T>;

namespace runtime {
builder::dyn_var<void*(int)> malloc = builder::with_name("malloc");
builder::dyn_var<void(void*)> free = builder::with_name("free");
builder::dyn_var<void(void*, void*, int)> memcpy = builder::with_name("memcpy");
builder::dyn_var<void(void)> start_time = builder::with_name("start_time");
builder::dyn_var<void(void)> end_time = builder::with_name("end_time");
builder::dyn_var<void(void)> printf = builder::with_name("printf");
builder::dyn_var<void(float*)> consume_tensor = builder::with_name("consume_tensor");
}
// A simple true at top boolean nd_var wrappable type
class false_top: public builder::nd_var_base {
public:
	typedef enum {
		F = 2,
		T = 1,
		Unspecified = 0,
	} level_t;
	
	typedef struct {
	    level_t level;
	    float threshold;    
	} value_type;
    value_type value;
	
	false_top(value_type def): value(def) {}
	false_top(): value({Unspecified, 0.0}) {}

	bool check(value_type e) {
	    if (value.level > e.level) 
	        return true;
	    if (value.level < e.level)
	        return false;
	    if (value.level == T)
	        return abs(value.threshold - e.threshold) < 0.001;
	    return true;
	}
	void merge(value_type e) {
	    // extra sanity check
	    if (check(e)) return;
	    if (value.level == T && e.level == T) {
	        value.level = F;
		    return;
	    }
	    value = e;
	}
};
|\clearpage|
struct tensor {
        int m_size;
        builder::dyn_var<float*> m_array;

        builder::static_var<bool> is_last_convolution = false;
        prophecy_var<false_top>* is_next_relu = nullptr;


        tensor(builder::dyn_var<float*> array, const int size): m_size(size), m_array(array) {}
        tensor(const int size): m_size(size), m_array(runtime::malloc(size * sizeof(float))) {}
        tensor(tensor& other): m_size(other.m_size), m_array(runtime::malloc(m_size * sizeof(float))) {
                runtime::memcpy(m_array, other.m_array, m_size * sizeof(float));
        }
        // move constructor cannibalizes the other tensor
        tensor(tensor&& other): m_size(other.m_size), m_array(other.m_array) {}

        void operator=(tensor& other) {
                if (m_size != other.m_size) {
                        runtime::free(m_array);
                        m_size = other.m_size;
                        m_array = runtime::malloc(m_size * sizeof(float));
                }
                runtime::memcpy(m_array, other.m_array, m_size * sizeof(float));
        }
        void operator=(tensor&& other) {
                runtime::free(m_array);
                m_size = other.m_size;
                m_array = other.m_array;
        }
};

tensor convolve(tensor &input, tensor &filter);
tensor convolve(tensor &input, tensor &filter) {
        tensor output (input.m_size);
        
        output.is_last_convolution = true;
        output.is_next_relu = new prophecy_var<false_top>();
        
        // simple convolution that follows wrap around behavior
        for (builder::dyn_var<int> i = 0; i < input.m_size; i++) {
                builder::dyn_var<float> sum = 0;
                for (builder::dyn_var<int> j = 0; j < filter.m_size; j++) {
                        sum += input.m_array[(i + j) % input.m_size] * filter.m_array[j];
                }
                if (output.is_next_relu->get().level == false_top::T) {
                    if (sum < output.is_next_relu->get().threshold) {
                        sum = 0;
                    }
                }
                output.m_array[i] = sum;
        }
        return output;
}
tensor relu(tensor &input, const float threshold);
tensor relu(tensor &input, const float threshold) {
        if (input.is_last_convolution) {
            input.is_last_convolution = false;
            input.is_next_relu->assert_requires_atleast({false_top::T, threshold});
            if (input.is_next_relu->get().level == false_top::T)
                return std::move(input);
        }
        for (builder::dyn_var<int> i = 0; i < input.m_size; i++) {
                if (input.m_array[i] >= threshold) {
                        input.m_array[i] = input.m_array[i];
                } else {
                        input.m_array[i] = 0;
                }
        }
        return std::move(input);
}
}

static builder::dyn_var<float*> benchmark(builder::dyn_var<float*> _input, const int size, 
            builder::dyn_var<float*> _weight, const int weight_size, builder::dyn_var<bool> small_t) {
        ml::tensor input(_input, size);
        ml::tensor weight(_weight, weight_size);

        ml::runtime::printf("Baseline:\\n");
        for (builder::dyn_var<int> i = 0; i < 10; i++) {
                ml::runtime::start_time();
                ml::tensor output_conv = ml::convolve(input, weight);
		        if (small_t) {
			        ml::tensor output = ml::relu(output_conv, 2.0);
			        ml::runtime::end_time();
			        ml::runtime::consume_tensor(output.m_array);
		        } else {
			        // Different threshold on a different branch
			        ml::tensor output = ml::relu(output_conv, 4.0);
			        ml::runtime::end_time();
			        ml::runtime::consume_tensor(output.m_array);
		        }
        }
	

        ml::runtime::printf("Optimized:\\n");
        for (builder::dyn_var<int> i = 0; i < 10; i++) {
                ml::runtime::start_time();
                ml::tensor output_conv = ml::convolve(input, weight);
                ml::tensor output = ml::relu(output_conv, 1.56);
                ml::runtime::end_time();
                ml::runtime::consume_tensor(output.m_array);
        }
        return 0;
}


int main(int argc, char* argv[]) {

        builder::builder_context ctx;
        ctx.run_rce = true;

        auto ast = ctx.extract_function_ast(benchmark, "benchmark", 10240, 21);
        std::cout << "#include \"ml_runtime.h\"" <<std::endl;
        block::c_code_generator::generate_code(ast, std::cout, 0);


        return 0;
}



\end{lstlisting}

\clearpage
\subsection{Implementation of union\_var<T>}
\label{sec:union}
This supplementary material provides the implementation of the \bldit{union_var<T>} type template containing a \bldit{dyn_var<T>} and \bldit{static_var<T>} component and a prophecy variable.
\\ \\ 
\begin{lstlisting}[language=buildit,escapechar=|]
#define MAX_ASSIGN_COUNT (16)
class binding_time: public builder::nd_var_base {
    typedef struct {
        bool is_dynamic = false;
        int assign_count = 0;
    } value_type;
    value_type value; 
    bool check (value_type e) {
        if (value.is_dynamic) return true;
        if (e.is_dynamic) return false;
        if (e.assign_count <= value.assign_count) return true;
        return false;
    }
    void merge (value_type e) {
        if (check(e)) return;
        value = e;
    }
};
template <typename T>
class union_var {
  dyn_var<T>             d_var = defer_init(); // defer the initialization after the prophecy_var check
  static_var<T>          s_var = defer_init(); // defer the initialization after the prophecy_var check
  // Default initialized to is_dynamic = false, assign_count = 0
  prophecy_var<binding_time>  needs_dynamic;
  // Keep track of the number of assignments to the variable
  int local_assign_count = 0;
  
  // Use prophecy variables to determine if this variable would be assigned a dynamic value in the future
  // and initialize the corresponding variable. The default creation of both the variables is suppressed 
  union_var () {
    if (needs_dynamic.get().is_dynamic) {
      d_var.init();
    } else {
      s_var.init();
    }
  }
  // If the initialization is dyn_var, no need to check the prophecy variable
  union_var(const dyn_var<T>& d) {
    d_var.init(d);
  }
  // Even if the initialization is a static value, future operations might make this variable dynamic
  union_var(const static_var<T>& d) {
    if (needs_dynamic.get().is_dynamic) {
      d_var.init(d);
    } else {
      s_var.init(d);
    }
  }
|\clearpage|
  // When the variable is assigned, update the prophecy variable values 
  union_var& operator= (union_var<T>& rhs) {
    if (needs_dynamic.get().is_dynamic) {
      if (rhs.needs_dynamic.get().is_dynamic) {
        d_var = rhs.d_var;
      } else {
        d_var = rhs.s_var;
      }
    } else {
      if (rhs.needs_dynamic.get().is_dynamic) {
        needs_dynamic.assert_requires_atleast({true, 0});
        // This is unreachable
      } else {
        // Check if too many assignments have been made
        local_assign_count++;
        if (local_assign_count > MAX_ASSIGN_COUNT) {
          needs_dynamic.assert_requires_atleast({true, 0});
        }
        // Move up the lattice
        needs_dynamic.assert_requires_atleast({false, local_assign_count});
        // If everything is okay, update the static value
        s_var = rhs.s_var;
      }
    }
  }
  // Rest of the operators and constructors
  ... 
}; 

    \end{lstlisting}

\end{document}